\documentclass[a4paper,fleqn]{cas-sc}

\usepackage[numbers,sort&compress]{natbib}
\usepackage{amsmath,amssymb}
\usepackage{array}
\usepackage[ruled]{algorithm2e}
\usepackage{bm}
\usepackage{booktabs}
\usepackage{graphicx}
\usepackage{iftex}
\usepackage[section]{placeins}
\usepackage{xspace}

\ifPDFTeX
  \newcommand{\TableTimesFont}{\fontfamily{ptm}\selectfont}
\else
  \usepackage{fontspec}
  \newfontfamily\TableTimesFont{Times New Roman}
\fi

\AtBeginEnvironment{table}{\TableTimesFont}
\AtBeginEnvironment{tabular}{\TableTimesFont}
\DontPrintSemicolon
\SetAlgoNoEnd
\SetAlFnt{\small}

\ExplSyntaxOn
\cs_set:Npn \__make_tbl_caption:nn #1#2
{
  \l_tbl_align_tl
  \skip_vertical:N \l_tbl_abovecap_skip
  {\parbox{ \dimexpr(\l_tbl_width_dim)}
    {\rightskip=0pt\TableTimesFont\small\textbf{\color{scolor}#1}\par#2\par\vskip4pt }}
  \skip_vertical:N \l_tbl_belowcap_skip
}
\cs_set:Npn \__make_fig_caption:nn #1#2
{
  \l_fig_align_tl
  \skip_vertical:N \l_fig_abovecap_skip
  \setbox\cascaptionbox=\hbox{%
     \TableTimesFont\small\textbf{\color{scolor}#1:}~#2}
  \ifdim\the\wd\cascaptionbox<\dim_use:N \l_fig_width_dim\relax
    \parbox{ \l_fig_width_dim }
      {\unskip\ignorespaces\hfil\TableTimesFont\small
       \textbf{\color{scolor}#1:}~#2\hfil\par }
  \else
    \parbox{ \l_fig_width_dim }
      {\rightskip=0pt\unskip\ignorespaces\TableTimesFont
       \small\textbf{\color{scolor}#1:}~#2\par }
  \fi
  \skip_vertical:N \l_fig_belowcap_skip
}
\ExplSyntaxOff

\def\tsc#1{\csdef{#1}{\textsc{\lowercase{#1}}\xspace}}
\tsc{LBM}

\newcommand{\Grad}{\nabla_{\!X}}
\newcommand{\Div}{\nabla_{\!X}\!\cdot}

\newcommand{\diag}{\operatorname{diag}}

\newcommand{\Id}{\bm I}
\newcommand{\bX}{\bm X}

\newcommand{\bx}{\bm x}
\newcommand{\bu}{\bm u}
\newcommand{\bv}{\bm v}
\newcommand{\bb}{\bm b}
\newcommand{\bF}{\bm F}

\newcommand{\bP}{\bm P}
\newcommand{\bT}{\bm T}
\newcommand{\bU}{\bm U}
\newcommand{\bB}{\bm B}
\newcommand{\bPhi}{\bm\Phi}
\newcommand{\bzero}{\bm 0}
\newcommand{\bn}{\bm n}
\newcommand{\bN}{\bm N}
\newcommand{\bt}{\bm t}
\newcommand{\bg}{\bm g}
\newcommand{\bh}{\bm h}
\newcommand{\be}{\bm e}
\newcommand{\bd}{\bm d}

\newcommand{\bw}{\bm w}
\newcommand{\dd}{\mathrm{d}}
\newcommand{\RR}{\mathbb R}

\begin{document}
\let\WriteBookmarks\relax
\renewcommand{\topfraction}{0.95}
\renewcommand{\bottomfraction}{0.85}
\renewcommand{\textfraction}{0.05}
\renewcommand{\floatpagefraction}{0.80}
\setcounter{topnumber}{3}
\setcounter{bottomnumber}{2}
\setcounter{totalnumber}{5}
\setlength{\floatsep}{10pt plus 2pt minus 2pt}
\setlength{\textfloatsep}{12pt plus 2pt minus 2pt}
\setlength{\intextsep}{10pt plus 2pt minus 2pt}

\shorttitle{Total-Lagrangian vectorial LBM for hyperelasticity with curved boundaries}
\shortauthors{J. Feng and X. Chu}

\title[mode=title]{Total-Lagrangian vectorial lattice Boltzmann method for finite-strain hyperelasticity with curved boundaries}

\author[1]{Jingsen Feng}

\author[1]{Xu Chu}
\cormark[1]
\ead{x.chu@exeter.ac.uk}

\affiliation[1]{
  organization={Department of Engineering, University of Exeter},
  city={Exeter},
  postcode={EX4 4QF},
  country={United Kingdom}
}

\cortext[1]{Corresponding author: Xu Chu, x.chu@exeter.ac.uk}

\begin{abstract}
Finite-strain hyperelasticity on curved embedded domains poses a geometric challenge for lattice Boltzmann methods.  After streaming across an embedded material surface, the missing population is recovered at the physical cut-link point, where the lattice direction, surface normal, and tangential deformation directions are generally distinct.  We develop a total-Lagrangian vectorial lattice Boltzmann method that resolves this geometric mismatch for two- and three-dimensional hyperelastic dynamics.  The continuum equations are written as a conservative first-order system for material velocity and deformation gradient.  Vector-valued populations are chosen so that their moments recover the state and the material-coordinate Piola fluxes, giving D2Q4\(\times\)6 and D3Q6\(\times\)12 schemes from one \(D\)-dimensional construction.  Curved boundaries are embedded by a level set and closed link by link through opposite-population moment identities, cut-link interpolation, and local geometric information at the boundary point.  The reconstruction is coupled to a compatibility projection that keeps the recovered displacement aligned with the evolved deformation gradient on embedded active-node graphs.  The resulting method extends the previous grid-aligned two-dimensional formulation to curved domains and three-dimensional lattices while retaining explicit collide--stream updates on Cartesian grids.  Benchmarks in two and three dimensions show agreement with exact finite-strain fields, nonlinear radial boundary-value problems, and finite-element references.
\end{abstract}


\maketitle

\section{Introduction}
\label{sec:introduction}

Large-deformation solid mechanics often involves domains whose boundaries are simple to describe and costly to fit with high-quality meshes.  Soft structures, biological tissues, elastomeric components, and architected materials commonly combine nonlinear stress response, transient motion, and curved material interfaces \citep{humphrey2003,jia2019architected,lengiewicz2020softcontact}.  Body-fitted Lagrangian finite elements remain a standard discretization for such problems \citep{bonetwood2008,belytschko2014}.  They represent the boundary directly and impose mechanical boundary conditions in a natural weak form.  Severe deformation can still degrade element quality and turn mesh generation or remeshing into an additional numerical problem.  Meshfree, particle-grid, immersed, and embedded-grid formulations address this cost by separating the material body from the computational background \citep{chen1996rkpm,wieckowski2004mpm,peskin2002,burman2015cutfem}.  Hyperelastic immersed-boundary methods show that finite-strain solid mechanics can be represented on such backgrounds, and recent stabilization studies for large-deformation incompressible elasticity highlight the care required when kinematics, stress, and boundary conditions are coupled through an immersed description \citep{boffi2008hyperelasticib,vadalaroth2020stabilization}.

The lattice Boltzmann method (LBM) offers a complementary route to background-grid computation, with an algorithmic appeal rooted in regular lattices, local updates, and natural suitability for parallel hardware \citep{mcnamara1988,qian1992,chen1998,succi2001,kruger2017}.  These features have made LBM a mature tool for fluid flow and transport, including recent applications to multiphase flow and tensor diffusion \citep{feng2026helmholtz,feng2026entropic}.  Curved-domain fluid calculations have also shown that geometric information can be incorporated on a fixed lattice without generating body-fitted meshes \citep{mei1999,bouzidi2001}.  This makes LBM appealing for large-deformation solids on embedded domains, provided the regular lattice update can be reconciled with finite-strain kinematics, nonlinear material response, and mechanical boundary conditions.

LBM formulations for solid mechanics have followed several paths.  For linear elastostatics, Yin et al. \citep{yin2016elastic} proposed displacement-distribution formulations for the Lam\'e equation, and Boolakee et al. developed second-order lattice Boltzmann schemes and boundary treatments for quasi-static linear elasticity on arbitrary two-dimensional domains \citep{boolakee2023static,boolakee2023bc}.  For elastodynamics, early solid and elastic-wave models were introduced by Marconi and Chopard \citep{marconi2003solid} and by O'Brien et al. \citep{obrien2012poisson}.  Later moment-chain formulations addressed tunable Poisson ratios, theoretical stability, surface waves, crack loading, and boundary conditions \citep{murthy2018elasticwaves,escande2020elasticwaves,schluter2018crack,faust2024bc}.  M\"uller et al. recently extended scalar-population moment-chain ideas to nonlinear finite-strain dynamics, with constitutive effects introduced through forcing terms and finite-difference gradient and divergence evaluations \citep{muller2025nonlinear}.  These studies established LBM as a possible solver for transient elastic response and exposed the structural difficulty of representing coupled elastic fluxes with scalar populations and standard hydrodynamic moment sets.

A separate line of development based on vectorial LBM is attractive here for structural reasons.  In this approach, each lattice direction carries a vector whose entries collect physical variables such as velocity and deformation measures.  Coupled mechanical quantities are therefore advanced within the same lattice update, preserving the local character of LBM and avoiding separate finite-difference reconstructions in the bulk.  It also gives a more systematic route from two-dimensional models to three-dimensional ones.  Additional mechanical components can be incorporated without changing the basic update pattern.  Related vectorial LBM and relaxation ideas have been developed for hyperbolic systems and finite-difference-based lattice Boltzmann schemes \citep{jin1995,graille2014,dubois2014vectorial,zhao2024}.  In solid mechanics, Boolakee et al. \citep{boolakee2025linear} used vectorial populations to construct a second-order LBM for linear elastodynamics.  Building on this route, Feng and Chu \citep{feng2026tlvlbm} introduced a total-Lagrangian vectorial LBM for finite-strain hyperelastic dynamics on grid-aligned two-dimensional domains.

The earlier total-Lagrangian vectorial formulation \citep{feng2026tlvlbm} was developed for grid-aligned two-dimensional domains, and curved embedded boundaries change the reconstruction problem.  Streaming leaves an incoming population undefined whenever a lattice link crosses the material surface.  The boundary point lies at a cut-link fraction between lattice nodes.  The missing population is associated with the lattice direction, the mechanical traction is defined with the surface normal, and the tangential part of the deformation gradient must be compatible with the displacement field on the boundary.  These directions coincide only in special grid-aligned cases.  For a generic curved surface, a consistent reconstruction has to distinguish the link flux, the physical traction, and the tangential deformation induced by the boundary displacement.

This paper extends the total-Lagrangian vectorial LBM to finite-strain hyperelasticity on curved embedded domains in two and three dimensions.  The bulk formulation is written once in dimension \(D\), yielding D2Q4\(\times\)6 in two dimensions and D3Q6\(\times\)12 in three dimensions.  The state consists of the material velocity and the full deformation gradient, and the equilibrium populations are constructed so that their moments recover the state together with the conservative Piola and kinematic fluxes.  The boundary is represented by a level set, which supplies active nodes, cut-link fractions, boundary points, and surface normals.  Opposite-population pair identities are then used row by row to impose the appropriate state or flux relation at each cut link, with link-wise interpolation transferring the relation from the physical boundary point to the active lattice node.  The same cut-link framework handles velocity and nominal-traction conditions, with the traction case constructing a compatible boundary deformation gradient from surface derivatives and a local nonlinear solve.  A compatibility projection aligns the recovered displacement and evolved deformation gradient during traction calculations.  The validation sequence isolates these ingredients through two-dimensional annulus and superellipse benchmarks, followed by three-dimensional ellipsoid, spherical-shell, and stretched-twisted tube benchmarks.

The remainder of the paper is arranged as follows.  Section~\ref{sec:tl-vectorial-lbm-2d3d} presents the total-Lagrangian first-order system and the D-dimensional vectorial lattice Boltzmann discretization.  Section~\ref{sec:curved-boundary-reconstruction} develops the curved cut-link reconstruction for Dirichlet and nominal-traction boundaries and describes the compatibility projection.  Section~\ref{sec:2d-curved-validation} examines the two-dimensional setting through annular and superelliptic embedded domains, separating curved velocity reconstruction, mixed traction loading, and spatially varying normal--tangent geometry.  Section~\ref{sec:3d-curved-validation} then carries the same ideas to the D3Q6\(\times\)12 lattice through ellipsoidal, spherical-shell, and stretched--twisted tube benchmarks.  Section~\ref{sec:conclusions-outlook} summarizes the main findings and outlines extensions toward broader material laws and multiphysics coupling.  A separate appendix records the grid-aligned D3Q6\(\times\)12 boundary formulas as a limiting case of the curved reconstruction.

\section{Total-Lagrangian vectorial lattice Boltzmann formulation in two and three dimensions}
\label{sec:tl-vectorial-lbm-2d3d}

This section identifies the continuum structure that the lattice Boltzmann method is constructed to reproduce.  We present the formulation in a unified notation for spatial dimensions \(D=2\) and \(D=3\).  The two-dimensional case corresponds to the D2Q4\(\times\)6 total-Lagrangian vectorial formulation used in the preceding finite-strain study \citep{feng2026tlvlbm}, where D2Q4\(\times\)6 denotes two spatial dimensions, four lattice directions, and six-component vector populations.  The three-dimensional case gives the analogous D3Q6\(\times\)12 extension, with six lattice directions and twelve-component vector populations.  In both dimensions, finite-strain elastodynamics is written as a first-order conservative system whose primary variables are the material velocity and the full deformation gradient.  The role of the lattice is then algebraic, with vector-valued populations chosen so that their moments recover both the state and the material-coordinate fluxes.

This viewpoint separates ingredients that are often intertwined in displacement-based discretizations.  The deformation-gradient evolution equation carries the kinematics, the pointwise stress map contains the constitutive nonlinearity, and the lattice discretization represents the conservative state and its material-coordinate fluxes through moments.  As a result, the same collide--stream update can accommodate different hyperelastic laws and extend from two to three dimensions without changing the basic algorithmic structure.

\subsection{Reference description, kinematics, and boundary conditions}
\label{subsec:kinematics-balance-d}

Let \(\Omega_0\subset\RR^D\), \(D\in\{2,3\}\), be the reference body.  Material coordinates are denoted by \(\bX=(X_1,\ldots,X_D)^T\), and the current position is
\begin{equation}
  \bx(\bX,t)=\bX+\bu(\bX,t).
  \label{eq:x-u-definition}
\end{equation}
The material velocity and deformation gradient are
\begin{equation}
  \bv=\partial_t\bu,
  \qquad
  \bF=\frac{\partial\bx}{\partial\bX}=\Id+\Grad\bu,
  \qquad
  F_{iA}=\delta_{iA}+\partial_Au_i,
  \label{eq:v-F-definition}
\end{equation}
where \(i=1,\ldots,D\) indexes spatial components, \(A=1,\ldots,D\) indexes material-coordinate directions, \(\partial_A=\partial/\partial X_A\), and \(\delta_{iA}\) is the Kronecker delta.  Repeated spatial and material indices are summed unless stated otherwise.  Admissible finite-strain states satisfy
\begin{equation}
  J=\det\bF>0.
  \label{eq:positive-jacobian}
\end{equation}
The positivity condition enforces orientation preservation in the continuum model, without which the Piola transformation and most compressible hyperelastic energies no longer define a physically meaningful stress.

A hyperelastic material is specified by a strain-energy density \(W(\bF)\) per unit reference volume \citep{ogden1997,holzapfel2000,bonetwood2008}.  The first Piola--Kirchhoff stress and material tangent are
\begin{equation}
  \bP(\bF)=\frac{\partial W}{\partial\bF},
  \qquad
  \mathbb C_{iAkB}(\bF)=\frac{\partial P_{iA}}{\partial F_{kB}},
  \qquad
  \delta P_{iA}=\mathbb C_{iAkB}\delta F_{kB}.
  \label{eq:P-C-general}
\end{equation}
All variables below are nondimensional.  With body acceleration \(\bb\), the total-Lagrangian equations are
\begin{equation}
  \partial_t v_i=\partial_A P_{iA}(\bF)+b_i,
  \qquad
  \partial_t F_{iA}=\partial_Av_i .
  \label{eq:momentum-F-general}
\end{equation}
The first relation is the material momentum balance in velocity form, and the second is obtained by differentiating \(F_{iA}=\delta_{iA}+\partial_Au_i\) in time.  Directly evolving \(\bF\) allows the bulk collision step to evaluate the stress from the local deformation gradient, without reconstructing strain from a separately evolved displacement field.

The initial conditions are prescribed as
\begin{equation}
  \bu(\bX,0)=\bu_0(\bX),
  \qquad
  \bv(\bX,0)=\bv_0(\bX),
  \qquad
  \bF(\bX,0)=\Id+\Grad\bu_0(\bX).
  \label{eq:initial-data-general}
\end{equation}
These fields define the initial first-order state.  The deformation gradient is initialized from the displacement field, which places \(\bF\) in the compatible class
\begin{equation}
  \partial_A F_{iB}-\partial_B F_{iA}=0
  \label{eq:F-compatibility-general}
\end{equation}
for smooth initial data.  Equation~\eqref{eq:momentum-F-general} preserves this relation at the continuum level.  The embedded-boundary algorithm uses the same compatibility relation when constructing Neumann boundary deformation gradients, where the quadrature displacement is replaced by a globally compatible field before the local traction inversion.

The reference boundary is decomposed into displacement and traction parts, \(\Gamma_D\) and \(\Gamma_N\).  On \(\Gamma_D\), the prescribed displacement is
\begin{equation}
  \bu=\bu_D .
  \label{eq:physical-dirichlet}
\end{equation}
The first-order lattice state evolves velocity, so the population reconstruction uses the boundary velocity
\begin{equation}
  \bv_D=\partial_t\bu_D .
  \label{eq:vD-from-uD}
\end{equation}
The displacement condition therefore enters the lattice through its time derivative, while the displacement itself is recovered by the velocity quadrature in the algorithm.

On \(\Gamma_N\), the natural boundary condition is written in reference form as
\begin{equation}
  \bP(\bF)\bN=\bar{\bT},
  \label{eq:nominal-traction-general}
\end{equation}
where \(\bN\) is the outward unit normal in the reference configuration.  The vector \(\bar{\bT}\) is a nominal traction per unit reference length in two dimensions and per unit reference area in three dimensions.  This is the traction conjugate to the first Piola--Kirchhoff stress in the total-Lagrangian balance.  If a load is specified as a spatial traction, it is converted through Nanson's relation, \(\bm\sigma\bn\,\dd a=\bP\bN\,\dd A_0\), where \(\bm\sigma\) is the Cauchy stress and \(\dd a\) and \(\dd A_0\) are current and reference surface measures, before being imposed in the population reconstruction \citep{ogden1997,holzapfel2000,bonetwood2008}.

\subsection{First-order conservative system}
\label{subsec:first-order-conservative-d}

The macroscopic state is chosen as
\begin{equation}
  \bU=\left(v_1,\ldots,v_D,F_{11},\ldots,F_{1D},F_{21},\ldots,F_{DD}\right)^T
  \in\RR^{D+D^2},
  \label{eq:state-general-d}
\end{equation}
with the deformation-gradient entries ordered row by row.  In this state, Eq.~\eqref{eq:momentum-F-general} becomes the conservative balance law
\begin{equation}
  \partial_t\bU+\partial_A\bPhi_A(\bU)=\bB,
  \qquad
  \bB=(b_1,\ldots,b_D,0,\ldots,0)^T .
  \label{eq:first-order-general-d}
\end{equation}
For each material-coordinate direction \(X_A\), the conservative system has a flux vector \(\bPhi_A(\bU)\in\RR^{D+D^2}\) with the same row structure as the state vector.  We denote by \(\Phi_A^{v_i}\) the entry associated with the velocity component \(v_i\), and by \(\Phi_A^{F_{iB}}\) the entry associated with the deformation-gradient component \(F_{iB}\).  These entries are
\begin{equation}
  \Phi_A^{v_i}(\bU)=-P_{iA}(\bF),
  \qquad
  \Phi_A^{F_{iB}}(\bU)=-\delta_{AB}v_i .
  \label{eq:flux-index-general-d}
\end{equation}
The first group of entries supplies the Piola-stress flux in the momentum rows.  The second group supplies the kinematic flux in the rows for \(\bF\), where the factor \(\delta_{AB}\) selects the column of \(\bF\) associated with the differentiated material coordinate.  The signs follow from placing the equations in the form \(\partial_t\bU+\partial_A\bPhi_A=\bB\).  Taking the divergence of Eq.~\eqref{eq:flux-index-general-d} gives \(-\partial_AP_{iA}\) in the velocity rows and \(-\partial_Av_i\) in the deformation-gradient rows, so Eq.~\eqref{eq:first-order-general-d} is exactly Eq.~\eqref{eq:momentum-F-general}.

In the component listings below, \(X\), \(Y\), and \(Z\) label the material-coordinate directions \(X_1\), \(X_2\), and \(X_3\), respectively.  For \(D=2\), the state and fluxes are
\begin{equation}
  \bU_{2D}=(v_1,v_2,F_{11},F_{12},F_{21},F_{22})^T,
  \label{eq:state-2d}
\end{equation}
\begin{equation}
  \bPhi_X=(-P_{11},-P_{21},-v_1,0,-v_2,0)^T,
  \qquad
  \bPhi_Y=(-P_{12},-P_{22},0,-v_1,0,-v_2)^T .
  \label{eq:flux-2d}
\end{equation}
For \(D=3\), the state is
\begin{equation}
  \bU_{3D}=
  (v_1,v_2,v_3,
  F_{11},F_{12},F_{13},
  F_{21},F_{22},F_{23},
  F_{31},F_{32},F_{33})^T,
  \label{eq:state-3d}
\end{equation}
and the three fluxes are
\begin{align}
  \bPhi_X&=(-P_{11},-P_{21},-P_{31},
  -v_1,0,0,-v_2,0,0,-v_3,0,0)^T, \notag\\
  \bPhi_Y&=(-P_{12},-P_{22},-P_{32},
  0,-v_1,0,0,-v_2,0,0,-v_3,0)^T, \notag\\
  \bPhi_Z&=(-P_{13},-P_{23},-P_{33},
  0,0,-v_1,0,0,-v_2,0,0,-v_3)^T .
  \label{eq:flux-3d}
\end{align}
The stress components in these fluxes are nonlinear functions of \(\bF\).  The flux Jacobians are therefore state dependent, but their action has a simple form:
\begin{equation}
  \left(\bPhi_A'(\bU)\delta\bU\right)^{v_i}
  =-\mathbb C_{iAkB}\delta F_{kB},
  \qquad
  \left(\bPhi_A'(\bU)\delta\bU\right)^{F_{iB}}
  =-\delta_{AB}\delta v_i .
  \label{eq:flux-jvp-general-d}
\end{equation}
Here the prime denotes the Fréchet derivative with respect to the state \(\bU\).  This Jacobian-vector form is the quantity needed by second-order initialization, wave-speed auditing, and local traction-boundary Newton iterations.  Dense \((D+D^2)\times(D+D^2)\) matrices are unnecessary.

For a unit material direction \(\bn\), the total-Lagrangian acoustic tensor is
\begin{equation}
  Q_{ik}(\bn;\bF)=n_A\mathbb C_{iAkB}(\bF)n_B .
  \label{eq:acoustic-tensor-general-d}
\end{equation}
If the material tangent is strongly elliptic, the eigenvalues of the matrix \(Q(\bn;\bF)\) are the squared tangent wave speeds in direction \(\bn\) after nondimensionalization.  The lattice speed \(c=\Delta x/\Delta t\) is chosen and audited against these local tangent speeds in the numerical studies.

\subsection{Moment space and equilibrium populations}
\label{subsec:vectorial-equilibrium-d}

Let \(\{\be_A\}_{A=1}^D\) be the standard Cartesian basis in the reference-coordinate space.  The lattice uses the axis-aligned velocity set
\begin{equation}
  \mathcal V_D=\{\pm\be_1,\ldots,\pm\be_D\},
  \label{eq:axis-velocity-set}
\end{equation}
with lattice spacing \(\Delta x\) in every coordinate direction and lattice velocity \(\bm c_q=c\be_q\), where \(c=\Delta x/\Delta t\) and \(\be_q=(e_{q1},\ldots,e_{qD})^T\in\mathcal V_D\).  Each direction carries a vector population
\begin{equation}
  \bm f_q(\bX,t)\in\RR^{D+D^2}.
  \label{eq:population-general}
\end{equation}
These populations are an algebraic lifting of the first-order continuum system.  They provide enough moment degrees of freedom to represent one state vector and \(D\) material-coordinate flux vectors.  The unused pair-sum degrees of freedom are distributed symmetrically among the coordinate directions, which avoids embedding a preferred axis into the equilibrium.

The equilibrium populations are required to satisfy
\begin{equation}
  \sum_q\bm f_q^{eq}=\bU,
  \qquad
  \sum_qc\,e_{qA}\bm f_q^{eq}=\bPhi_A(\bU),
  \qquad A=1,\ldots,D,
  \label{eq:moment-constraints-general}
\end{equation}
with the additional symmetric pair-sum closure
\begin{equation}
  \bm f_{+A}^{eq}+\bm f_{-A}^{eq}=\frac{1}{D}\bU,
  \qquad A=1,\ldots,D .
  \label{eq:pair-sum-general}
\end{equation}
Here \(+A\) and \(-A\) denote the lattice directions \(\be_A\) and \(-\be_A\), respectively.
Solving these linear constraints gives
\begin{equation}
  \bm f_q^{eq}(\bU)=\frac{1}{2D}
  \left[
  \bU+\frac{D}{c}e_{qA}\bPhi_A(\bU)
  \right] .
  \label{eq:feq-general-d}
\end{equation}
The identities \(\sum_qe_{qA}=0\) and \(\sum_qe_{qA}e_{qB}=2\delta_{AB}\) verify Eq.~\eqref{eq:moment-constraints-general}.  For \(D=2\), \(\be_q\in\{\pm\be_1,\pm\be_2\}\) and Eq.~\eqref{eq:feq-general-d} becomes the D2Q4\(\times\)6 equilibrium
\begin{equation}
  \bm f_q^{eq}=\frac14
  \left[
  \bU+\frac{2}{c}e_{qA}\bPhi_A(\bU)
  \right],
  \qquad \be_q\in\{\pm\be_1,\pm\be_2\}.
  \label{eq:feq-2d}
\end{equation}
For \(D=3\), \(\be_q\in\{\pm\be_1,\pm\be_2,\pm\be_3\}\) and Eq.~\eqref{eq:feq-general-d} gives the D3Q6\(\times\)12 equilibrium
\begin{equation}
  \bm f_q^{eq}=\frac16
  \left[
  \bU+\frac{3}{c}e_{qA}\bPhi_A(\bU)
  \right],
  \qquad \be_q\in\{\pm\be_1,\pm\be_2,\pm\be_3\}.
  \label{eq:feq-3d}
\end{equation}
Finite-strain material response enters Eq.~\eqref{eq:feq-general-d} only through \(\bPhi_A(\bU)\), hence only through \(\bP(\bF)\).  The moment-matching algebra is the same for all hyperelastic laws admitted by the local stress map.

The derivative of the equilibrium is
\begin{equation}
  \left(\bm f_q^{eq}\right)'(\bU)\bm Z
  =
  \frac{1}{2D}
  \left[
  \bm Z+\frac{D}{c}e_{qA}\bPhi_A'(\bU)\bm Z
  \right].
  \label{eq:feq-derivative-general}
\end{equation}
Taking moments of Eq.~\eqref{eq:feq-derivative-general} recovers \(\bm Z\) and \(\bPhi_A'(\bU)\bm Z\).  The same algebra that represents the nonlinear fluxes also represents their tangent action.

\subsection{Collision, streaming, forcing, and initialization}
\label{subsec:collision-streaming-d}

The state is recovered from populations with a half-step source correction,
\begin{equation}
  \bU^{num}(\bX,t)=\sum_q\bm f_q(\bX,t)+\frac{\Delta t}{2}\bB(\bX,t).
  \label{eq:state-recovery-general}
\end{equation}
According to the ordering in Eq.~\eqref{eq:state-general-d}, \(\bU^{num}\) is split into the velocity block and the row-wise deformation-gradient block.  The latter is reshaped into \(\bF\), the constitutive law evaluates \(\bP(\bF)\), and the fluxes and equilibria are then computed locally.  With relaxation parameter \(0<\omega<2\), the Bhatnagar--Gross--Krook (BGK) collision step \citep{qian1992,succi2001,kruger2017} is
\begin{equation}
  \bm f_q^{*}
  =
  \bm f_q
  +\omega\left(\bm f_q^{eq}(\bU^{num})-\bm f_q\right)
  +\Delta t(2-\omega)w_q\bB .
  \label{eq:collision-general-d}
\end{equation}
The source weights are
\begin{equation}
  w_q=\frac{1}{4D},
  \qquad
  \sum_qw_q=\frac12,
  \qquad
  \sum_qe_{qA}w_q=0 .
  \label{eq:source-weights-general}
\end{equation}
Thus \(w_q=1/8\) for D2Q4 and \(w_q=1/12\) for D3Q6.  The source contributes to the recovered state while leaving the flux moments centered.  The post-collision populations stream to neighboring active lattice nodes,
\begin{equation}
  \bm f_q(\bX+c\be_q\Delta t,t+\Delta t)=\bm f_q^{*}(\bX,t).
  \label{eq:streaming-general-d}
\end{equation}

For smooth prescribed initial fields, let \(\bU_0=\bU(\bX,0)\) and \(\bB_0=\bB(\bX,0)\).  A second-order population initialization follows from expanding the equilibrium distribution one half time step backward along its lattice characteristic:
\begin{equation}
  \bm f_q(\bX,0)
  =
  \bm f_q^{eq}(\bU_0)
  -\frac{\Delta t}{2}
  D_q\bm f_q^{eq}(\bU_0),
  \qquad
  D_q=\partial_t+c\,e_{qA}\partial_A .
  \label{eq:init-general-d}
\end{equation}
The missing time derivative is supplied by the continuum balance law,
\begin{equation}
  \partial_t\bU_0=\bB_0-\bPhi_A'(\bU_0)\partial_A\bU_0,
  \label{eq:Ut-general-init}
\end{equation}
and Eq.~\eqref{eq:feq-derivative-general} evaluates \(D_q\bm f_q^{eq}\).  Analytical derivatives are used for manufactured solutions.  Second-order finite differences are used when only discrete benchmark initial data are available.  When the computation starts from a spatially uniform undeformed state, such as \(\bv_0=\bzero\) and \(\bF_0=\Id\), equilibrium initialization is sufficient since the characteristic correction vanishes and no leading kinetic layer is introduced.

The displacement is a dependent field used for diagnostics and for compatibility operations at embedded boundaries.  It is recovered by trapezoidal velocity quadrature,
\begin{equation}
  \bu^n=\bu^{*,n}+\frac{\Delta t}{2}\bv^n,
  \qquad
  \bu^{*,n+1}=\bu^n+\frac{\Delta t}{2}\bv^n .
  \label{eq:u-recovery-general}
\end{equation}
where \(\bu^{*,n}\) is the stored history variable for the trapezoidal update.
The stress in the dynamics is always \(\bP(\bF)\).  If a spatial stress is required for post-processing, the Cauchy stress is recovered as
\begin{equation}
  \bm\sigma=J^{-1}\bP\bF^T .
  \label{eq:cauchy-recovery-general}
\end{equation}

\section{Arbitrary curved-boundary reconstruction}
\label{sec:curved-boundary-reconstruction}

The bulk update described above is local and independent of boundary geometry.  Each active node undergoes the same moment-matched collision, and interior streaming only transfers post-collision populations along lattice links.  Curved boundaries enter when a streamed population would leave the material body.  The boundary treatment can then be posed as the reconstruction of the missing incoming population on each cut link \citep{mei1999,bouzidi2001}.  In this formulation, geometry modifies the streaming step, while the constitutive update remains the same local map \(\bF\mapsto\bP\) used in the interior.

The reconstruction uses three ingredients.  The level-set description identifies active nodes, cut fractions, and boundary normals.  Opposite-population pair identities determine whether a given row should impose a state component or a link-flux component.  A link interpolation then transfers the half-way pair relation to the physical boundary intersection.  The same construction applies to Dirichlet and nominal-traction Neumann conditions in two and three dimensions.

\subsection{Level-set geometry and cut links}
\label{subsec:cut-link-geometry}

The reference body is embedded in the Cartesian lattice through a level-set function \citep{sethian1999}
\begin{equation}
  \Omega_0=\{\bX:\phi(\bX)\le0\},
  \qquad
  \Gamma=\{\bX:\phi(\bX)=0\}.
  \label{eq:level-set-domain}
\end{equation}
The sign convention places active lattice nodes inside the material body, so a node \(\bX_f\) is active when \(\phi(\bX_f)\le0\).  For a lattice direction \(\be_q\), the link from \(\bX_f\) to \(\bX_f+\Delta x\,\be_q\) is an interior link if the neighboring node is also active.  If the neighbor lies outside the body, the link is a cut link.  We denote the outgoing direction on such a link by \(\be=\be_{q_{out}}\).  The physical boundary point lies between the active node and the outside node and is written as
\begin{equation}
  \bX_b=\bX_f+\eta\,\Delta x\,\be,
  \qquad
  \phi(\bX_b)=0,
  \qquad
  0<\eta<1 .
  \label{eq:cut-link-point}
\end{equation}
Here \(\eta\) is the fraction of the lattice link from the active node to the boundary.  The value \(\eta=1/2\) corresponds to a half-way boundary, while smaller or larger values place the boundary closer to the active node or the outside node.  In the implementation, \(\eta\) is found by bisection of \(\phi\) along the segment.  The outward unit normal is either supplied analytically or computed from the level set as
\begin{equation}
  \bN_b=\frac{\Grad\phi(\bX_b)}{\|\Grad\phi(\bX_b)\|} .
  \label{eq:level-set-normal}
\end{equation}
This formula is consistent with the convention \(\phi\le0\) inside the body, for which \(\Grad\phi\) points outward at a smooth boundary.  Boundary identifiers attached to cut links allow different portions of the same implicit surface to carry different boundary conditions.

The streaming step identifies the population to be reconstructed.  After collision at \(\bX_f\), the population moving in the outgoing direction \(\be\) is known, but normal streaming would send it to the outside node.  The population that should arrive at \(\bX_f\) from outside the body is not stored anywhere.  This missing population is associated with the incoming lattice direction
\begin{equation}
  \bd=-\be=-\be_{q_{out}},
  \qquad
  q_{in}=\operatorname{opp}(q_{out}).
  \label{eq:incoming-direction}
\end{equation}
where \(\operatorname{opp}(q)\) denotes the index of the lattice direction opposite to \(q\).
The reconstruction therefore determines the incoming population in direction \(\bd\) from the known opposite population and the prescribed boundary condition at \(\bX_b\).  The geometric normal \(\bN_b\) and the lattice direction \(\bd\) are kept separate throughout the construction.  The normal enters the physical traction \(\bP\bN_b\), and the lattice direction enters the population moments through the link flux \(d_A\bPhi_A\).  On curved surfaces these two directions are generally not parallel.

\subsection{Boundary pair relation and cut-link interpolation}
\label{subsec:pair-identities}
\label{subsec:cut-link-interpolation}

The population reconstruction starts from the two moments carried by an opposite lattice pair.  For directions \((\bd,-\bd)\), the equilibrium in Eq.~\eqref{eq:feq-general-d} gives
\begin{equation}
  \bm f_d^{eq}+\bm f_{-d}^{eq}=\frac{1}{D}\bU,
  \qquad
  \bm f_d^{eq}-\bm f_{-d}^{eq}=\frac{1}{c}d_A\bPhi_A(\bU).
  \label{eq:pair-identities-general}
\end{equation}
The pair sum represents the state, while the pair difference represents the flux in the lattice-link direction.  A prescribed state component is therefore imposed through the pair sum, and a prescribed link-flux component is imposed through the pair difference.  For each cut link, these choices are collected in a local relation at the boundary point,
\begin{equation}
  \bm f_d^b=\bm D_b\bm f_{-d}^b+\bm S_b,
  \label{eq:boundary-pair-relation}
\end{equation}
where \(\bm D_b\) selects the pair parity row by row and \(\bm S_b\) contains the boundary values entering the selected pair sums or pair differences.  For a half-way boundary, Eq.~\eqref{eq:boundary-pair-relation} directly gives the missing incoming population.  For a general cut location, the relation is imposed at \(\bX_b\) and interpolated back to the active lattice node.  With \(\bm f^*\) denoting post-collision populations, the Bouzidi--Firdaouss--Lallemand-type interpolation \citep{bouzidi2001} reads, for \(\eta\ge1/2\),
\begin{equation}
  \bm f_d(\bX_f,t+\Delta t)
  =
  \frac{\bm D_b\bm f_{-d}^{*}(\bX_f,t)+\bm S_b}{2\eta}
  +
  \left(1-\frac{1}{2\eta}\right)
  \bm f_d^{*}(\bX_f,t),
  \label{eq:bfl-eta-ge-half}
\end{equation}
and, for \(\eta<1/2\),
\begin{equation}
  \bm f_d(\bX_f,t+\Delta t)
  =
  \bm D_b
  \left[
  2\eta\,\bm f_{-d}^{*}(\bX_f,t)
  +(1-2\eta)\bm f_{-d}^{*}(\bX_f-\Delta x\,\be,t)
  \right]
  +\bm S_b .
  \label{eq:bfl-eta-lt-half}
\end{equation}
If \(\bX_f-\Delta x\,\be\) is not an active node, the upstream value is replaced by the local value \(\bm f_{-d}^{*}(\bX_f,t)\).  This fallback sacrifices the second upstream sample only where the local geometry does not provide it.  The half-way rule is recovered by setting \(\eta=1/2\).

\subsection{Velocity Dirichlet boundary}
\label{subsec:curved-dirichlet}

On a displacement boundary, the physical prescription enters the first-order system through the boundary velocity \(\bv_b=\bv_D(\bX_b,t+\tfrac12\Delta t)\).  The velocity rows impose this state value through the pair sum in Eq.~\eqref{eq:pair-identities-general}.  The deformation-gradient rows impose the corresponding kinematic link flux, which follows from Eq.~\eqref{eq:flux-index-general-d} as \(d_A\Phi_A^{F_{iB}}=-d_Bv_i\).  Substitution into the pair identities gives
\begin{equation}
  f_d^{v_i,b}
  =
  -f_{-d}^{v_i,b}
  +\frac{v_{b,i}}{D},
  \qquad
  f_d^{F_{iB},b}
  =
  f_{-d}^{F_{iB},b}
  -\frac{d_Bv_{b,i}}{c}.
  \label{eq:dirichlet-rows-curved}
\end{equation}
The superscripts \(v_i\) and \(F_{iB}\) denote the corresponding rows of the vector population.
In the compact boundary relation Eq.~\eqref{eq:boundary-pair-relation}, this corresponds to
\begin{equation}
  \bm D_D=\diag(-I_D,I_{D^2}),
  \qquad
  S_D^{v_i}=\frac{v_{b,i}}{D},
  \qquad
  S_D^{F_{iB}}=-\frac{d_Bv_{b,i}}{c}.
  \label{eq:dirichlet-compact-curved}
\end{equation}
where \(I_m\) denotes the \(m\times m\) identity matrix.
The rule enforces the velocity trace and the associated kinematic flux along the missing population link.

\subsection{Nominal-traction Neumann boundary}
\label{subsec:curved-neumann}

On a traction boundary, the prescribed condition is \(\bP(\bF_b)\bN_b=\bar{\bT}_b\), where \(\bar{\bT}_b=\bar{\bT}(\bX_b,t+\tfrac12\Delta t)\) is the nominal traction at the cut point and \(\bF_b\) is the deformation gradient assigned there.  Once \(\bF_b\) is known, \(\bP_b=\bP(\bF_b)\) supplies the stress flux in the velocity rows, while the deformation-gradient rows impose the boundary state.  The pair relation becomes
\begin{equation}
  f_d^{v_i,b}
  =
  f_{-d}^{v_i,b}
  -\frac{d_AP_{b,iA}}{c},
  \qquad
  f_d^{F_{iB},b}
  =
  -f_{-d}^{F_{iB},b}
  +\frac{F_{b,iB}}{D}.
  \label{eq:neumann-rows-curved}
\end{equation}
In the compact form of Eq.~\eqref{eq:boundary-pair-relation},
\begin{equation}
  \bm D_N=\diag(I_D,-I_{D^2}),
  \qquad
  S_N^{v_i}=-\frac{d_AP_{b,iA}}{c},
  \qquad
  S_N^{F_{iB}}=\frac{F_{b,iB}}{D}.
  \label{eq:neumann-compact-curved}
\end{equation}

The boundary deformation gradient \(\bF_b\) is constructed by fixing its tangential images from displacement compatibility and determining its normal image from the traction equation.  For a tangent direction \(\bt_\alpha\) at \(\bX_b\), compatibility gives
\begin{equation}
  \bF_b\bt_\alpha
  =
  \bt_\alpha+\Grad\bu(\bX_b,t)\bt_\alpha,
  \qquad \alpha=1,\ldots,D-1 .
  \label{eq:tangent-image-compatible}
\end{equation}
The displacement \(\bu\) here is the field recovered by velocity quadrature.  The evolved \(\bF\) still supplies the bulk stress, while the recovered displacement supplies only the surface derivatives needed to define the tangential part of \(\bF_b\).

The tangential derivatives are evaluated by a compact weighted least-squares fit over a local stencil \(\mathcal N_b\) of active nodes around the cut point.  In two dimensions, with normalized local coordinates
\begin{equation}
  \bm\xi_m=\frac{\bX_m-\bX_b}{\Delta x},
  \qquad \bX_m\in\mathcal N_b,
  \label{eq:lsq-coordinates-2d}
\end{equation}
each displacement component is fitted to
\begin{equation}
  p(\xi_1,\xi_2)
  =
  a_0+a_1\xi_1+a_2\xi_2
  +\frac12a_3\xi_1^2+a_4\xi_1\xi_2+\frac12a_5\xi_2^2 .
  \label{eq:quadratic-lsq-2d}
\end{equation}
If \(\bm A_{\rm lsq}\) is the polynomial matrix and \(\bm W_{\rm lsq}\) is the diagonal least-squares weight matrix, the derivative along a unit tangent \(\bt=(t_1,t_2)^T\) is recovered with
\begin{equation}
\begin{aligned}
  \bw_t^T
  &=
  \begin{bmatrix}
  0 & t_1/\Delta x & t_2/\Delta x & 0 & 0 & 0
  \end{bmatrix}
  (\bm A_{\rm lsq}^T\bm W_{\rm lsq}\bm A_{\rm lsq})^{-1}\bm A_{\rm lsq}^T\bm W_{\rm lsq},
  \\
  \Grad u_i(\bX_b,t)\bt
  &\approx
  \sum_{\bX_m\in\mathcal N_b} w_{t,m}u_i(\bX_m,t).
\end{aligned}
  \label{eq:tangent-derivative-2d}
\end{equation}
Thus the two-dimensional tangent image is \(\bg=\bt+\sum_{\bX_m\in\mathcal N_b} w_{t,m}\bu(\bX_m,t)\).  In three dimensions, an orthonormal surface frame \(\{\bN_b,\bt_1,\bt_2\}\) is used, with \(\bt_\alpha\cdot\bN_b=0\) and \(\bt_1\cdot\bt_2=0\).  Two least-squares derivative rows give
\begin{equation}
  \bg_\alpha
  =
  \bt_\alpha
  +
  \sum_{\bX_m\in\mathcal N_b} w_{\alpha,m}\bu(\bX_m,t),
  \qquad \alpha=1,2 .
  \label{eq:tangent-image-3d-lsq}
\end{equation}

With the tangent images fixed, the remaining unknown is the normal image.  In two dimensions,
\begin{equation}
  \bF_b=\bh\otimes\bN_b+\bg\otimes\bt,
  \qquad
  \bh=\bF_b\bN_b,
  \label{eq:Fb-decomp-2d}
\end{equation}
and \(\bh\in\RR^2\) is determined from
\begin{equation}
  R_i(\bh)
  =
  P_{iA}\!\left(\bh\otimes\bN_b+\bg\otimes\bt\right)N_{b,A}
  -\bar T_{b,i}
  =0,
  \qquad i=1,2 .
  \label{eq:neumann-newton-2d}
\end{equation}
In three dimensions,
\begin{equation}
  \bF_b
  =
  \bh\otimes\bN_b
  +\bg_1\otimes\bt_1
  +\bg_2\otimes\bt_2,
  \qquad
  \bh=\bF_b\bN_b,
  \label{eq:Fb-decomp-3d}
\end{equation}
and \(\bh\in\RR^3\) solves
\begin{equation}
  R_i(\bh)
  =
  P_{iA}\!\left(
  \bh\otimes\bN_b+
  \bg_1\otimes\bt_1+
  \bg_2\otimes\bt_2
  \right)N_{b,A}
  -\bar T_{b,i}
  =0,
  \qquad i=1,2,3 .
  \label{eq:neumann-newton-3d}
\end{equation}
A small Newton iteration with a finite-difference residual Jacobian is sufficient because the solve is local to a single cut link.  Trial states are accepted only when \(\det\bF_b\) remains positive.  The least-squares fit supplies only surface derivatives, and the constitutive law remains the same pointwise map used in the bulk.

\subsection{Compatibility projection}
\label{subsec:compatibility-projection}

The continuum relation \(\bF=\Id+\Grad\bu\) ties the evolved deformation gradient to the quadrature displacement.  In any embedded-domain step that contains nominal-traction Neumann cut links, the compatibility operation is performed globally on the active-node graph.  It produces a compatible displacement \(\bu^c\) whose graph differences match the currently evolved \(\bF-\Id\), writes this displacement back to the quadrature variables, and repairs the deformation-gradient populations so that the subsequent bulk evolution starts from a compatible first-order state.  The projection is therefore an in-step state update in the reported Neumann calculations.  After collision, the displacement projection is applied again before cut-link reconstruction, so the surface derivatives used in Eqs.~\eqref{eq:tangent-image-compatible}--\eqref{eq:tangent-image-3d-lsq} are evaluated from the projected displacement.  Dirichlet-only calculations do not use this global compatibility operation.

For neighboring active nodes \(\bX_\ell\) and \(\bX_m\), the projection uses the edge relation
\begin{equation}
  \bu^c(\bX_m)-\bu^c(\bX_\ell)
  \approx
  \frac12\left[(\bF_m-\Id)+(\bF_\ell-\Id)\right](\bX_m-\bX_\ell).
  \label{eq:projection-edge-relation}
\end{equation}
The resulting sparse normal equations are relaxed by damped Jacobi iterations on the active-node graph.  A single gauge condition, or prescribed displacement values when they are used as projection anchors, fixes the null mode of \(\bu^c\).  This nearest-neighbor form remains compatible with parallel active-node storage.
After the projection solve, the stored displacement variables are reset as
\begin{equation}
  \bu_m\leftarrow\bu_m^c,
  \qquad
  \bu_m^*\leftarrow\bu_m^c-\frac{\Delta t}{2}\bv_m .
  \label{eq:projection-displacement-reset}
\end{equation}
The compatible deformation gradient used for repair is
\begin{equation}
  \bF_m^c=\Id+\nabla_{\Delta x}\bu^c(\bX_m),
  \label{eq:projection-compatible-F}
\end{equation}
where \(\nabla_{\Delta x}\) denotes the active-node finite-difference gradient, with one-sided differences where a centered stencil is cut by the embedded boundary.  The deformation-gradient state after repair is
\begin{equation}
  \bF_m^r
  =
  (1-\theta_F)\bF_m+\theta_F\bF_m^c,
  \qquad 0<\theta_F\le1 .
  \label{eq:projection-F-repair-weight}
\end{equation}
The reported Neumann calculations use the full repair \(\theta_F=1\), so that \(\bF_m^r=\bF_m^c\) on the active lattice.  The populations are translated by the corresponding equilibrium difference,
\begin{equation}
  \bm f_{q,m}
  \leftarrow
  \bm f_{q,m}
  +\bm f_q^{eq}(\bv_m,\bF_m^r)
  -\bm f_q^{eq}(\bv_m,\bF_m),
  \qquad
  \bF_m\leftarrow\bF_m^r .
  \label{eq:projection-population-repair}
\end{equation}
This update preserves the kinetic non-equilibrium part while changing the hydrodynamic deformation-gradient moment.

\subsection{One update on an embedded curved domain}
\label{subsec:one-step-curved}

At time \(t^n=n\Delta t\), the stored fields on the active lattice \(\Omega_{\Delta x}=\{\bX_m:\phi(\bX_m)\le0\}\) are \(\{\bm f_{q,m}^{\,n},\bu_m^{*,n}\}_{\bX_m\in\Omega_{\Delta x}}\).
The subscript \(m\) labels an active lattice node, \(q\) labels a lattice direction, and the superscript \(n\) labels the time level.  Thus \(\bm f_{q,m}^{\,n}\) is the population in direction \(q\) stored at node \(\bX_m\), while \(m+q\) denotes the neighboring node located at \(\bX_m+\Delta x\,\be_q\).
Interior and cut links are
\[
\begin{aligned}
  \mathcal I_{\Delta x}
  =
  \left\{(m,q):\bX_m\in\Omega_{\Delta x},\;\bX_m+\Delta x\,\be_q\in\Omega_{\Delta x}\right\},
  \\
  \mathcal C_{\Delta x}
  =
  \left\{(m,q):\bX_m\in\Omega_{\Delta x},\;\bX_m+\Delta x\,\be_q\notin\Omega_{\Delta x}\right\},
  \\
  \mathcal C_{\Delta x}^{N}
  =
  \left\{(m,q)\in\mathcal C_{\Delta x}:\bX_b(m,q)\in\Gamma_N\right\}.
\end{aligned}
\]
Here \(\mathcal I_{\Delta x}\) is the set of interior streaming links whose neighboring node remains in the active lattice, while \(\mathcal C_{\Delta x}\) is the set of cut links whose streaming target lies outside the embedded domain.  The subset \(\mathcal C_{\Delta x}^{N}\) contains the cut links whose boundary intersection lies on a nominal-traction Neumann boundary.  Thus \(\mathcal C_{\Delta x}^{N}\ne\emptyset\) means that the current embedded lattice contains at least one Neumann cut link.  This is the algorithmic trigger for the global compatibility operation: in the reported calculations, the cut-link interpolation is always Eqs.~\eqref{eq:bfl-eta-ge-half}--\eqref{eq:bfl-eta-lt-half}, and whenever \(\mathcal C_{\Delta x}^N\ne\emptyset\), the global compatibility projection and repair in Section~\ref{subsec:compatibility-projection} are part of the time step.  Algorithm~\ref{alg:one-step-curved} gives the resulting update.

\begin{algorithm}[H]
\footnotesize
\caption{One update on an embedded curved domain}
\label{alg:one-step-curved}
\KwIn{\(\{\bm f_{q,m}^{\,n},\bu_m^{*,n}\}_{\bX_m\in\Omega_{\Delta x}}\), body forces \(\bB^n\) and \(\bB^{n+1}\), cut-link geometry \((\eta,\bX_b,\bN_b)\), and boundary values on \(\Gamma_D\cup\Gamma_N\)}
\KwOut{\(\{\bm f_{q,m}^{\,n+1},\bu_m^{*,n+1}\}_{\bX_m\in\Omega_{\Delta x}}\)}

\ForEach{\(\bX_m\in\Omega_{\Delta x}\)}{
  Recover \(\bU_m^{num,n}\gets\sum_q\bm f_{q,m}^{\,n}+(\Delta t/2)\bB_m^n\), with \(\bU_m^{num,n}=(\bv_m^n,\bF_m^n)\)\;
  Recover \(\bu_m^n\gets\bu_m^{*,n}+(\Delta t/2)\bv_m^n\)\;
}

\If{\(\mathcal C_{\Delta x}^{N}\ne\emptyset\) for the pre-collision compatibility repair}{
  Solve globally on the active-node graph for \(\bu^{c,n}\) satisfying \(\bu_m^{c,n}-\bu_\ell^{c,n}\approx\frac12[(\bF_m^n-\Id)+(\bF_\ell^n-\Id)](\bX_m-\bX_\ell)\) on neighboring active-node edges; cf. Eq.~\eqref{eq:projection-edge-relation}\;
  Reset \(\bu_m^n\gets\bu_m^{c,n}\) and \(\bu_m^{*,n}\gets\bu_m^{c,n}-(\Delta t/2)\bv_m^n\) for every \(\bX_m\in\Omega_{\Delta x}\); cf. Eq.~\eqref{eq:projection-displacement-reset}\;
  Set \(\bF_m^{c,n}\gets\Id+\nabla_{\Delta x}\bu^{c,n}(\bX_m)\) for every \(\bX_m\in\Omega_{\Delta x}\); cf. Eq.~\eqref{eq:projection-compatible-F}\;
  Set \(\bF_m^{r,n}\gets(1-\theta_F)\bF_m^n+\theta_F\bF_m^{c,n}\) for every \(\bX_m\in\Omega_{\Delta x}\), with \(\theta_F=1\) in the reported calculations; cf. Eq.~\eqref{eq:projection-F-repair-weight}\;
  For every \(\bX_m\in\Omega_{\Delta x}\) and every \(q\), repair \(\bm f_{q,m}^{\,n}\gets\bm f_{q,m}^{\,n}+\bm f_q^{eq}(\bv_m^n,\bF_m^{r,n})-\bm f_q^{eq}(\bv_m^n,\bF_m^n)\), then set \(\bF_m^n\gets\bF_m^{r,n}\) and \(\bU_m^{num,n}\gets(\bv_m^n,\bF_m^n)\); cf. Eq.~\eqref{eq:projection-population-repair}\;
}

\ForEach{\(\bX_m\in\Omega_{\Delta x}\)}{
  Evaluate \(\bP_m^n\gets\bP(\bF_m^n)\), \(\bPhi_{A,m}^n\gets\bPhi_A(\bU_m^{num,n})\), and \(\bm f_{q,m}^{eq,n}\gets(2D)^{-1}\left[\bU_m^{num,n}+(D/c)e_{qA}\bPhi_{A,m}^n\right]\)\;
  \ForEach{lattice direction \(q\)}{
    Collide locally, \(\bm f_{q,m}^{*,n}\gets\bm f_{q,m}^{\,n}+\omega(\bm f_{q,m}^{eq,n}-\bm f_{q,m}^{\,n})+\Delta t(2-\omega)w_q\bB_m^n\)\;
  }
  Advance the displacement history, \(\bu_m^{*,n+1}\gets\bu_m^n+(\Delta t/2)\bv_m^n\)\;
}

\If{\(\mathcal C_{\Delta x}^{N}\ne\emptyset\) for the boundary displacement projection}{
  Solve globally on the active-node graph for \(\widehat{\bu}^{c,n}\) satisfying \(\widehat{\bu}_m^{c,n}-\widehat{\bu}_\ell^{c,n}\approx\frac12[(\bF_m^n-\Id)+(\bF_\ell^n-\Id)](\bX_m-\bX_\ell)\) on neighboring active-node edges; cf. Eq.~\eqref{eq:projection-edge-relation}\;
  Reset \(\bu_m^n\gets\widehat{\bu}_m^{c,n}\) and \(\bu_m^{*,n+1}\gets\widehat{\bu}_m^{c,n}-(\Delta t/2)\bv_m^n\) for every \(\bX_m\in\Omega_{\Delta x}\); cf. Eq.~\eqref{eq:projection-displacement-reset}\;
}

\ForEach{\((m,q)\in\mathcal I_{\Delta x}\)}{
  Stream to the neighbor, \(\bm f_{q,m+q}^{\,n+1}\gets\bm f_{q,m}^{*,n}\), where \(\bX_{m+q}=\bX_m+\Delta x\,\be_q\)\;
}

\ForEach{\((m,q)\in\mathcal C_{\Delta x}\)}{
  Set \(\be\gets\be_q\), \(\bd\gets-\be\), and \(\bX_b\gets\bX_m+\eta\,\Delta x\,\be\)\;
  \uIf{\(\bX_b\in\Gamma_D\)}{
    Set \(\bv_b\gets\bv_D(\bX_b,t^n+\Delta t/2)\), \(\bm D_b\gets\bm D_D\), \(S_b^{v_i}\gets v_{b,i}/D\), and \(S_b^{F_{iB}}\gets-d_Bv_{b,i}/c\)\;
  }
  \ElseIf{\(\bX_b\in\Gamma_N\)}{
    Compute \(\bg_\alpha\gets\bt_\alpha+\nabla_{\Delta x}\widehat{\bu}^{c,n}(\bX_b)\bt_\alpha\), \(\alpha=1,\ldots,D-1\)\;
    Solve \(\bP(\bF_b(\bh))\bN_b=\bar{\bT}_b\), with \(\bF_b(\bh)=\bh\otimes\bN_b+\sum_{\alpha=1}^{D-1}\bg_\alpha\otimes\bt_\alpha\)\;
    Set \(\bP_b\gets\bP(\bF_b)\), \(\bm D_b\gets\bm D_N\), \(S_b^{v_i}\gets-d_AP_{b,iA}/c\), and \(S_b^{F_{iB}}\gets F_{b,iB}/D\)\;
  }
  \uIf{\(\eta\ge1/2\)}{
    Set \(\bm f_{d,m}^{\,n+1}\gets\left(\bm D_b\bm f_{-d,m}^{*,n}+\bm S_b\right)/(2\eta)+\left(1-1/(2\eta)\right)\bm f_{d,m}^{*,n}\); cf. Eq.~\eqref{eq:bfl-eta-ge-half}\;
  }
  \Else{
    Set \(m^-\) by \(\bX_{m^-}=\bX_m-\Delta x\,\be\) if this node lies in \(\Omega_{\Delta x}\), and set \(m^-=m\) otherwise\;
    Set \(\bm f_{d,m}^{\,n+1}\gets\bm D_b\left[2\eta\,\bm f_{-d,m}^{*,n}+(1-2\eta)\bm f_{-d,m^-}^{*,n}\right]+\bm S_b\); cf. Eq.~\eqref{eq:bfl-eta-lt-half}\;
  }
}

\ForEach{\(\bX_m\in\Omega_{\Delta x}\)}{
  Recover \(\bU_m^{num,n+1}\gets\sum_q\bm f_{q,m}^{\,n+1}+(\Delta t/2)\bB_m^{n+1}\), with \(\bU_m^{num,n+1}=(\bv_m^{n+1},\bF_m^{n+1})\)\;
  Evaluate diagnostics \(\bP_m^{n+1}\gets\bP(\bF_m^{n+1})\), \(J_m^{n+1}\gets\det\bF_m^{n+1}\), and \(\bm\sigma_m^{n+1}\gets(J_m^{n+1})^{-1}\bP_m^{n+1}(\bF_m^{n+1})^T\)\;
}
\end{algorithm}

This ordering reflects the implementation used in the Neumann benchmarks.  The existence of Neumann cut links triggers a global compatibility projection over the active lattice.  Before collision, the projected displacement is written back and the deformation-gradient populations are repaired, so the compatibility operation affects the following bulk evolution.  After collision, a second displacement reset supplies the compatible field used by the Neumann tangential reconstruction.  Dirichlet cut links still use only the prescribed velocity and the kinematic link flux in their local reconstruction.  Cut-link reconstruction supplies the missing incoming populations through the pair identities and the interpolation formulas in Eqs.~\eqref{eq:bfl-eta-ge-half}--\eqref{eq:bfl-eta-lt-half}.

\section{Two-dimensional curved-boundary validation and benchmarks}
\label{sec:2d-curved-validation}

The preceding section gives the algorithmic ingredients of the embedded-boundary update, including population reconstruction on cut links, local Dirichlet and Neumann closures, and the compatibility projection used when traction boundaries are present.  The next step is to evaluate these ingredients in geometries where the exact boundary is not aligned with the lattice.  The validation therefore begins in two dimensions, where analytic and finite-element references allow the individual effects of curved Dirichlet data, curved tractions, and mixed boundary identifiers to be examined before moving to the fully three-dimensional cases.

The two-dimensional validation is organized to separate the error mechanisms introduced by the embedded-boundary treatment.  The bulk D2Q4\(\times\)6 update, the constitutive law, and the nondimensional scaling are kept fixed throughout the sequence, so that changes in accuracy can be attributed primarily to the cut-link reconstruction, the local traction inversion, and the global compatibility projection activated by Neumann boundaries.  The cases are arranged so that each step adds one source of geometric or boundary-condition difficulty.  The affine annulus is deliberately simple in the bulk.  Its exact deformation gradient is uniform, so the case mainly probes how accurately the curved boundary supplies velocity Dirichlet data.  The second case keeps the annular geometry but replaces the outer boundary condition by a radial nominal traction.  This introduces a mixed Dirichlet--Neumann relaxation problem with a one-dimensional static limit, while the numerical solution is still obtained through the transient ramp-and-relax LBM dynamics.  The final two-dimensional case then removes circular symmetry.  The rotated superellipse shell retains mixed boundary types, but its nonuniform wall thickness and changing normals make the cut-link fractions and local tangent frames vary along the interface.

Although the target states in the last two cases are steady equilibria, they are obtained here by advancing the first-order lattice system from an initially relaxed configuration under smooth loading.  For a lattice Boltzmann method, this distinction matters because the comparison concerns the full transient collide--stream dynamics with curved reconstruction and checks whether it reaches the relaxed finite-strain state, instead of substituting a static elliptic solve.

\subsection{Material, scaling, and error measures}
\label{subsec:2d-material-scaling}

All validation cases use the hyperelastic update described in Section~\ref{sec:tl-vectorial-lbm-2d3d}.  The same material model is used in two and three dimensions so that geometry and boundary reconstruction are examined with the constitutive response held fixed.  The material is the compressible neo-Hookean solid with strain-energy density \citep{ogden1997,holzapfel2000,bonetwood2008}
\begin{equation}
  W(\bF)
  =
  \frac{\mu}{2}\left(\|\bF\|_F^2-D\right)
  -\left(\mu+\frac{\lambda}{2}\right)\ln J
  +\frac{\lambda}{4}\left(J^2-1\right),
  \qquad
  J=\det\bF>0.
  \label{eq:validation-nh-energy}
\end{equation}
Here \(D=2\) in the present section and \(D=3\) in Section~\ref{sec:3d-curved-validation}, and \(\|\cdot\|_F\) is the Frobenius norm.  The nondimensional material parameters are the shear modulus \(\mu=1\) and Poisson ratio \(\nu=0.20\), giving Lamé's first parameter \(\lambda=2/3\).  The stress entering the evolution is \(\bP=\partial W/\partial\bF\), and the Cauchy stress plotted below is recovered by Eq.~\eqref{eq:cauchy-recovery-general}.

The plotted dimensional values use one physical interpretation of the nondimensional calculations.  The length scale is \(L_0=0.1~{\rm m}\), the stress scale is \(S_0=10~{\rm MPa}\), and the reference density is \(\rho_0=1000~{\rm kg\,m^{-3}}\), which give \(V_0=\sqrt{S_0/\rho_0}=100~{\rm m\,s^{-1}}\) and \(T_0=L_0\sqrt{\rho_0/S_0}=1~{\rm ms}\).  Coordinates and displacements are reported in centimetres, stresses and nominal tractions in MPa, and time in milliseconds.  The deformation gradient, \(J\), relative errors, and convergence orders remain nondimensional.  With this scaling, \(\mu=10~{\rm MPa}\), \(\lambda=6.667~{\rm MPa}\), Young's modulus is \(E=24~{\rm MPa}\), and the bulk modulus is \(K=13.333~{\rm MPa}\).

For any sampled field \(Q\), relative errors are evaluated on the active lattice nodes as
\begin{equation}
  E_{L2}(Q)=\frac{\|Q_{\rm LBM}-Q_{\rm ref}\|_2}{\|Q_{\rm ref}\|_2},
  \qquad
  E_{L2}(\bF-\Id)=
  \frac{\|\bF_{\rm LBM}-\bF_{\rm ref}\|_2}{\|\bF_{\rm ref}-\Id\|_2}.
  \label{eq:validation-error-measures}
\end{equation}
Here \(\|\cdot\|_2\) is the discrete Euclidean norm over the comparison nodes.  For the affine solution, the stress error is measured against the exact Cauchy stress.  For the traction benchmarks, finite-element fields are sampled at the lattice nodes and used as the reference.  The displacement error measures the accumulated velocity quadrature and boundary-state reconstruction.  Errors in \(\bF\) and \(\sigma\) are more stringent because they involve the evolved deformation gradient and the nonlinear stress map; they are therefore more sensitive to boundary-local derivative reconstruction and to the constitutive inversion used in the traction condition.

The same numerical settings are used in the two- and three-dimensional validation cases unless stated otherwise in the individual benchmark descriptions.  All lattice calculations use the nondimensional lattice speed \(c=10\), so that \(\Delta t=\Delta x/10\) on each grid.  The traction and mixed-boundary benchmarks use the BGK relaxation parameter \(\omega=1.8\).  For the ramp-and-relax calculations, the source term in Eq.~\eqref{eq:collision-general-d} is used as velocity-proportional damping, \(\bB_{\rm damp}=(-\gamma v_1,\ldots,-\gamma v_D,0,\ldots,0)^T\), with \(\gamma=10\).  This term dissipates elastic-wave transients before the terminal comparison and leaves the static hyperelastic target defined by Eq.~\eqref{eq:validation-nh-energy} unchanged as the velocity decays.  Whenever Neumann cut links are present, the compatibility projection described in Section~\ref{subsec:compatibility-projection} is applied every 20 LBM steps.  The affine Dirichlet annulus and rotated ellipsoid have only prescribed-velocity boundaries and therefore use the local Dirichlet reconstruction without the global compatibility projection; their relaxation parameters are \(\omega=2.0\) and \(\omega=1.8\), respectively.

\subsection{Affine Dirichlet annulus}
\label{subsec:2d-affine-annulus}

The affine annulus is the most restrictive Dirichlet benchmark because the exact deformation gradient and stress are uniform in the material.  Consequently, the continuum solution contains no interior stress-divergence contribution, and the remaining spatial error reflects how accurately the curved boundary supplies the incoming populations.  The annular geometry also exposes the reconstruction to two smooth closed interfaces with opposite normal orientations while retaining an exact finite-strain solution.  The reference body is
\begin{equation}
  \Omega_0
  =
  \left\{\bX:\ R_i\le \|\bX-\bX_c\|\le R_o\right\},
  \qquad
  \bX_c=(0.5,0.5)^T,\quad
  R_i=0.18,\quad
  R_o=0.42 .
  \label{eq:affine-annulus-domain}
\end{equation}
This is a concentric circular shell embedded in the unit square.  The material occupies only the ring between the inner circle and the outer circle; the circular hole is excluded from the active lattice.  Both circular boundaries are smooth cut-link surfaces and both are assigned the same analytical velocity Dirichlet data.  The off-diagonal entries of the imposed affine map generate shear and rotation-like components, so the benchmark is less degenerate than a purely radial expansion.  In physical units the inner and outer radii are \(1.8~{\rm cm}\) and \(4.2~{\rm cm}\).  The prescribed motion is affine,
\begin{equation}
  \bu_{\rm ex}(\bX,t)
  =
  \frac{t}{T}
  \bm A(\bX-\bX_c),
  \qquad
  \bv_{\rm ex}(\bX,t)=\frac{1}{T}\bm A(\bX-\bX_c),
  \qquad
  0\le t\le T,
  \label{eq:affine-annulus-solution}
\end{equation}
with
\begin{equation}
  T=0.2,\qquad
  \bm A=
  \begin{pmatrix}
    0.35 & 0.18\\
   -0.10 & 0.24
  \end{pmatrix}.
  \label{eq:affine-annulus-A}
\end{equation}
The terminal deformation gradient is uniform,
\begin{equation}
  \bF_*=\Id+\bm A=
  \begin{pmatrix}
    1.35 & 0.18\\
   -0.10 & 1.24
  \end{pmatrix},
  \qquad
  J_*=1.692 .
  \label{eq:affine-annulus-F}
\end{equation}
Since \(\bP(\bF(t))\) is spatially constant, \(\Div\bP=\bzero\).  The exact bulk dynamics are therefore generated entirely by the boundary velocity in Eq.~\eqref{eq:affine-annulus-solution}, without body acceleration.  The exact Cauchy stress follows from the neo-Hookean stress and Eq.~\eqref{eq:cauchy-recovery-general}; at \(t=T\),
\begin{equation}
  \bm\sigma_{\rm ex}
  =
  \begin{pmatrix}
    8.7226 & 0.5213\\
    0.5213 & 6.9064
  \end{pmatrix}
  {\rm MPa}.
  \label{eq:affine-annulus-stress}
\end{equation}

\begin{figure}[pos=htbp]
\centering
\includegraphics[width=0.98\textwidth]{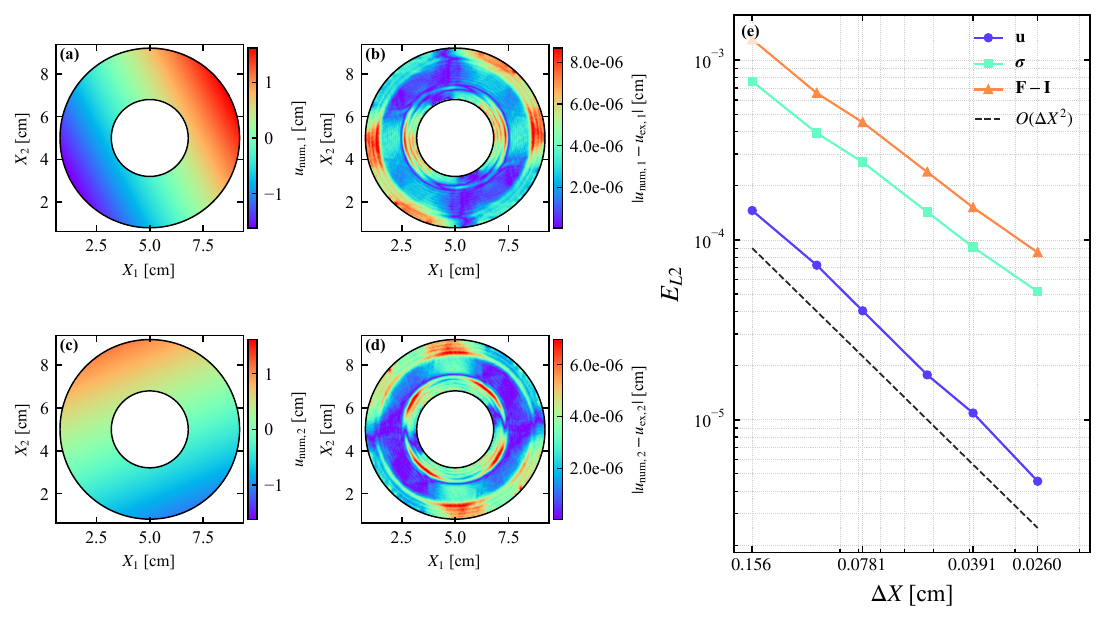}
\caption{Affine Dirichlet annulus. Panels (a,c) show the numerical displacement components at \(T=0.2~{\rm ms}\) on the \(N=384\) grid. Panels (b,d) show the absolute component errors relative to Eq.~\eqref{eq:affine-annulus-solution}. Panel (e) reports grid convergence for displacement, Cauchy stress, and \(\bF-\Id\).}
\label{fig:2d-affine-annulus}
\end{figure}

Figure~\ref{fig:2d-affine-annulus} shows the terminal displacement components and componentwise errors on the \(N=384\) grid, together with the grid refinement from \(N=64\) to \(384\).  The finest grid has \(\Delta x=0.0260~{\rm cm}\), \(66692\) active nodes, and \(1840\) cut links.  The maximum displacement magnitude is \(1.656~{\rm cm}\).  At this resolution the relative errors are \(4.55\times10^{-6}\) for displacement, \(8.51\times10^{-5}\) for \(\bF-\Id\), and \(5.16\times10^{-5}\) for Cauchy stress.  The displacement error follows the second-order reference trend across the refinement range.  The deformation-gradient and stress errors have larger constants, as expected for derivative and constitutive quantities evaluated near cut links, yet they decrease consistently under grid refinement.  This case therefore verifies that the curved Dirichlet reconstruction preserves an exact finite-strain affine mode to high accuracy on a non-boundary-fitted annular lattice.

\FloatBarrier

\subsection{Radial traction annulus}
\label{subsec:2d-radial-annulus}

The second benchmark activates the traction branch of the reconstruction while retaining a reference structure that can be interpreted mechanically.  The same concentric annulus is used with different boundary types on the two circular components.  The inner cavity is clamped, while the outer surface is loaded by a nominal radial traction.  The inner circle is the boundary of the hole, so its outward normal points toward the centre of the annulus.  The outer circle bounds the exterior of the shell and has the usual radial outward normal.  This configuration checks both boundary identifiers and normal-orientation conventions.  With \(R=\|\bX-\bX_c\|\), the inner boundary is clamped,
\begin{equation}
  \bu=\bzero \qquad \text{on } R=R_i,
  \label{eq:radial-annulus-inner-bc}
\end{equation}
and the outer boundary carries a radial load
\begin{equation}
  \bP\bN=p(t)\be_R,
  \qquad
  p(t)=p_0s(t),
  \qquad
  p_0=1,
  \qquad
  R=R_o,
  \label{eq:radial-annulus-outer-bc}
\end{equation}
where \(\be_R=(\bX-\bX_c)/\|\bX-\bX_c\|\).  The smooth loading factor is
\begin{equation}
  s(t)=
  \begin{cases}
    \sin^2\!\left(\dfrac{\pi t}{2T_r}\right), & 0\le t<T_r,\\[5pt]
    1, & t\ge T_r,
  \end{cases}
  \qquad
  T_r=2 .
  \label{eq:sine2-ramp}
\end{equation}
In the physical scaling above, the final traction amplitude is \(10~{\rm MPa}\), the ramp time is \(2~{\rm ms}\), and the run time is \(5~{\rm ms}\).

The radial specialization of the static continuum problem gives a useful reference form \citep{ogden1997,holzapfel2000}.  With
\begin{equation}
  \bx(\bX)=\bX_c+r(R)\be_R,
  \qquad
  R=\|\bX-\bX_c\|,
  \label{eq:radial-map}
\end{equation}
the principal stretches are
\begin{equation}
  \lambda_r=\frac{\dd r}{\dd R},
  \qquad
  \lambda_\theta=\frac{r}{R},
  \qquad
  J=\lambda_r\lambda_\theta .
  \label{eq:radial-stretches-2d}
\end{equation}
The neo-Hookean principal nominal stresses are
\begin{align}
  P_r
  &=
  \mu\left(\lambda_r-\lambda_r^{-1}\right)
  +\frac{\lambda}{2}\left(J^2-1\right)\lambda_r^{-1},
  \notag\\
  P_\theta
  &=
  \mu\left(\lambda_\theta-\lambda_\theta^{-1}\right)
  +\frac{\lambda}{2}\left(J^2-1\right)\lambda_\theta^{-1}.
  \label{eq:radial-principal-stresses-2d}
\end{align}
The radial equilibrium equation is
\begin{equation}
  \frac{\dd P_r}{\dd R}+\frac{P_r-P_\theta}{R}=0,
  \qquad
  r(R_i)=R_i,
  \qquad
  P_r(R_o)=p_0 .
  \label{eq:radial-equilibrium-2d}
\end{equation}
Equation~\eqref{eq:radial-equilibrium-2d} identifies the static equilibrium selected by the boundary conditions.  The LBM calculation reaches this state through the transient first-order dynamics.  The traction is ramped, elastic waves are damped, and the solution is sampled after relaxation.  The finite-element method (FEM) reference \citep{bonetwood2008} is advanced with the same material, geometry, final traction, sine-squared ramp, damping coefficient, time step, and final time as the lattice run, using the boundary-fitted quadratic mesh described below.  The comparison therefore uses an independent discretization under the same loading and relaxation protocol.

\begin{figure}[pos=htbp]
\centering
\includegraphics[width=0.98\textwidth]{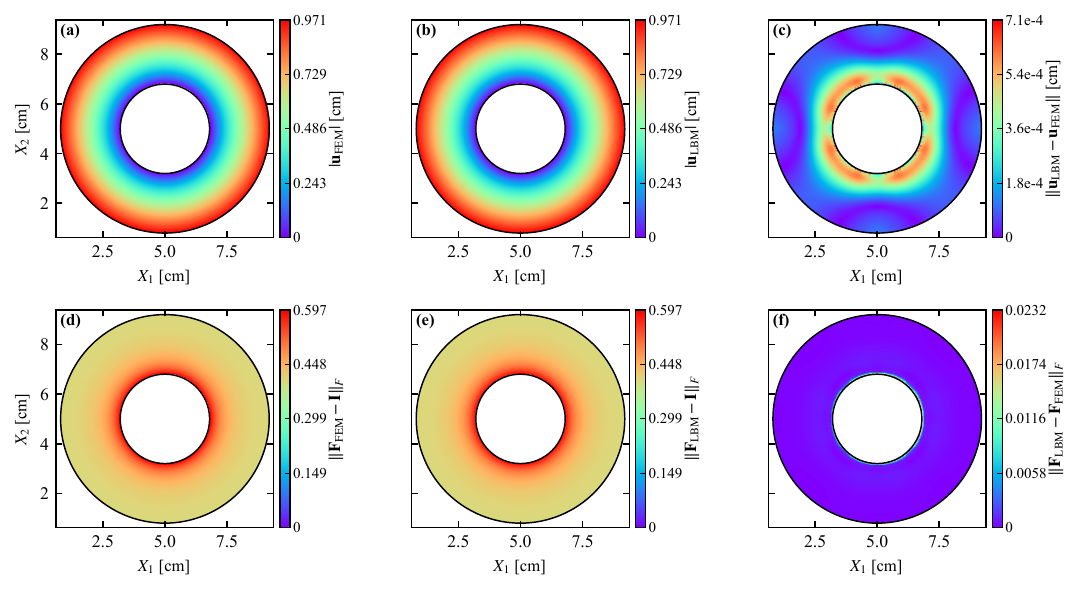}
\caption{Radial traction annulus at \(t=5~{\rm ms}\). The first row compares the finite-element and LBM displacement magnitudes and their pointwise difference. The second row compares \(\|\bF-\Id\|_F\) and \(\|\bF_{\rm LBM}-\bF_{\rm FEM}\|_F\).}
\label{fig:2d-radial-annulus-field}
\end{figure}

Figure~\ref{fig:2d-radial-annulus-field} compares displacement magnitude and deformation-gradient increment on the \(N=196\) lattice.  The LBM grid has \(\Delta x=1/196\), corresponding to \(0.0510~{\rm cm}\), with \(17396\) active nodes and \(936\) cut links.  The finite-element reference uses quadratic Lagrange elements on a boundary-fitted second-order annulus mesh with characteristic size \(h=1/196\), so the reference mesh size matches the LBM lattice spacing.  The two fields are sampled on \(17388\) common active nodes.  At \(t=5~{\rm ms}\), the maximum displacement magnitude is \(0.9715~{\rm cm}\) in the LBM solution and \(0.9714~{\rm cm}\) in the finite-element solution.  The relative errors are \(3.86\times10^{-4}\) for displacement, \(9.18\times10^{-4}\) for \(\bF\), and \(3.84\times10^{-3}\) for \(\bF-\Id\).  The terminal Jacobian remains in the range \(1.581\le J\le1.639\) over the comparison nodes.  The close agreement in displacement amplitude indicates that the ramped transient has relaxed to the same finite deformation as the independent reference.  The larger relative error in \(\bF-\Id\) reflects the small-denominator normalization and the sensitivity of gradient quantities to the Neumann reconstruction adjacent to the two circular boundaries.

\begin{figure}[pos=htbp]
\centering
\includegraphics[width=0.98\textwidth]{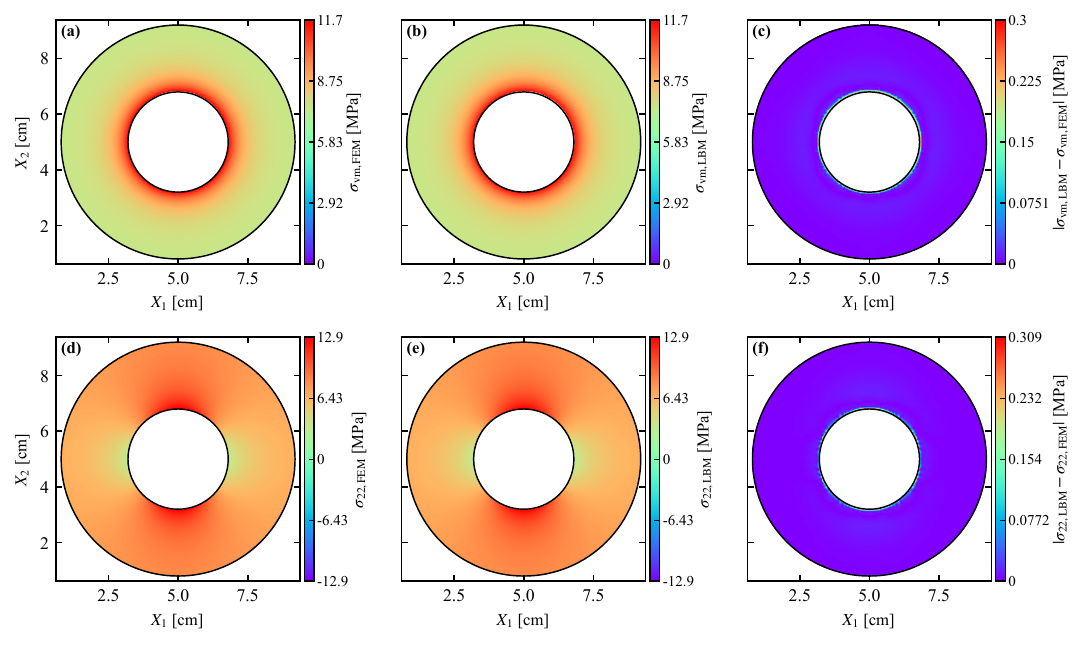}
\caption{Cauchy-stress comparison for the radial traction annulus. The panels show the finite-element and LBM von Mises stress, the absolute von Mises difference, the \(\sigma_{22}\) component, and its absolute difference.}
\label{fig:2d-radial-annulus-stress}
\end{figure}

Figure~\ref{fig:2d-radial-annulus-stress} gives the corresponding Cauchy-stress comparison.  The sampled von Mises stress reaches \(11.67~{\rm MPa}\) in the finite-element reference and \(11.48~{\rm MPa}\) in the LBM calculation.  The \(\sigma_{22}\) component spans \(3.30\) to \(12.86~{\rm MPa}\) for FEM and \(3.23\) to \(12.61~{\rm MPa}\) for LBM.  The maximum absolute von Mises difference is \(0.30~{\rm MPa}\).  The stress error is concentrated near the two circular interfaces, consistent with the deformation-gradient error in Fig.~\ref{fig:2d-radial-annulus-field} because the Cauchy stress is recovered directly from \(\bF\).  These boundary-local differences are associated with the Neumann reconstruction, where the traction inversion, surface derivative, and cut-link interpolation enter the update.  Away from the embedded boundaries, the stress follows from the same local constitutive evaluation used in the grid-aligned formulation and the LBM and FEM fields remain nearly indistinguishable.

\FloatBarrier

\subsection{Mixed-boundary superellipse shell}
\label{subsec:2d-superellipse-shell}

The final two-dimensional benchmark removes the radial symmetry used in the preceding annulus case.  The shell is bounded by two rotated superellipses, producing a nonuniform wall thickness and a distribution of boundary normals that is decoupled from the Cartesian lattice.  This geometry exercises the boundary treatment in the regime for which the local normal--tangent construction was introduced, combining mixed boundary identifiers, spatially varying curvature, and the absence of a one-dimensional reference reduction.  For a superellipse with centre \(\bX_c\), semi-axes \(a,b\), rotation angle \(\theta\), and exponent \(p=4\), define local coordinates
\begin{equation}
  \begin{pmatrix}
    \xi\\ \zeta
  \end{pmatrix}
  =
  \begin{pmatrix}
    \cos\theta & \sin\theta\\
   -\sin\theta & \cos\theta
  \end{pmatrix}
  (\bX-\bX_c),
  \qquad
  \phi(\bX)
  =
  \left|\frac{\xi}{a}\right|^p
  +
  \left|\frac{\zeta}{b}\right|^p
  -1 .
  \label{eq:superellipse-level-set}
\end{equation}
Let \(\phi_o\) and \(\phi_i\) denote this level-set function with the outer and inner parameter sets, respectively.  The shell is represented by
\begin{equation}
  \Omega_0
  =
  \left\{\bX:\max\left(\phi_o(\bX),-\phi_i(\bX)\right)\le0\right\}.
  \label{eq:superellipse-shell-domain}
\end{equation}
Both superellipses are centred at \((0.5,0.5)^T\).  The outer boundary has \(a_o=0.42\), \(b_o=0.30\), and \(\theta_o=20^\circ\); the inner boundary has \(a_i=0.16\), \(b_i=0.10\), and \(\theta_i=-15^\circ\).  The two rotations are different, so the wall thickness varies around the shell and the boundary normals are not aligned with the Cartesian grid.  Boundary identifiers separate the inner and outer superellipses during cut-link reconstruction.  In physical units these semi-axes are \(4.2~{\rm cm}\), \(3.0~{\rm cm}\), \(1.6~{\rm cm}\), and \(1.0~{\rm cm}\).

The inner boundary is clamped by the velocity Dirichlet rule.  The outer boundary carries a vertical nominal traction, so the load direction is fixed in the reference frame while the boundary normal varies along the superellipse.  This separates the physical traction vector from the lattice-link direction in the reconstruction.  The imposed traction is
\begin{equation}
  \bP\bN=s(t)(0,0.125)^T,
  \label{eq:superellipse-traction}
\end{equation}
with the same ramp \(s(t)\) as Eq.~\eqref{eq:sine2-ramp}.  The final physical traction is \((0,1.25)^T~{\rm MPa}\).  The ramp-and-relax LBM simulation is advanced to \(t=20~{\rm ms}\), which corresponds to \(76800\) collide--stream steps on the \(N=384\) lattice.  A dynamic finite-element reference advanced with the same time step would require the same number of global updates on a boundary-fitted quadratic mesh, making the reference calculation much more expensive than a terminal equilibrium solve.  The LBM terminal field is therefore compared with an independent static finite-element equilibrium computed for the same material, geometry, inner Dirichlet constraint, and terminal nominal traction.  This comparison keeps the finite-element calculation focused on the final equilibrium state, while the lattice calculation resolves the full transient ramp-and-relax path.  The finite-element reference uses quadratic Lagrange elements on a boundary-fitted second-order superellipse mesh with characteristic size \(h=1/384\), equal to the LBM lattice spacing \(\Delta x\) and corresponding to \(0.0260~{\rm cm}\).  The finite-element fields are sampled at the same active lattice nodes used for the LBM error evaluation.

\begin{figure}[pos=htbp]
\centering
\includegraphics[width=0.98\textwidth]{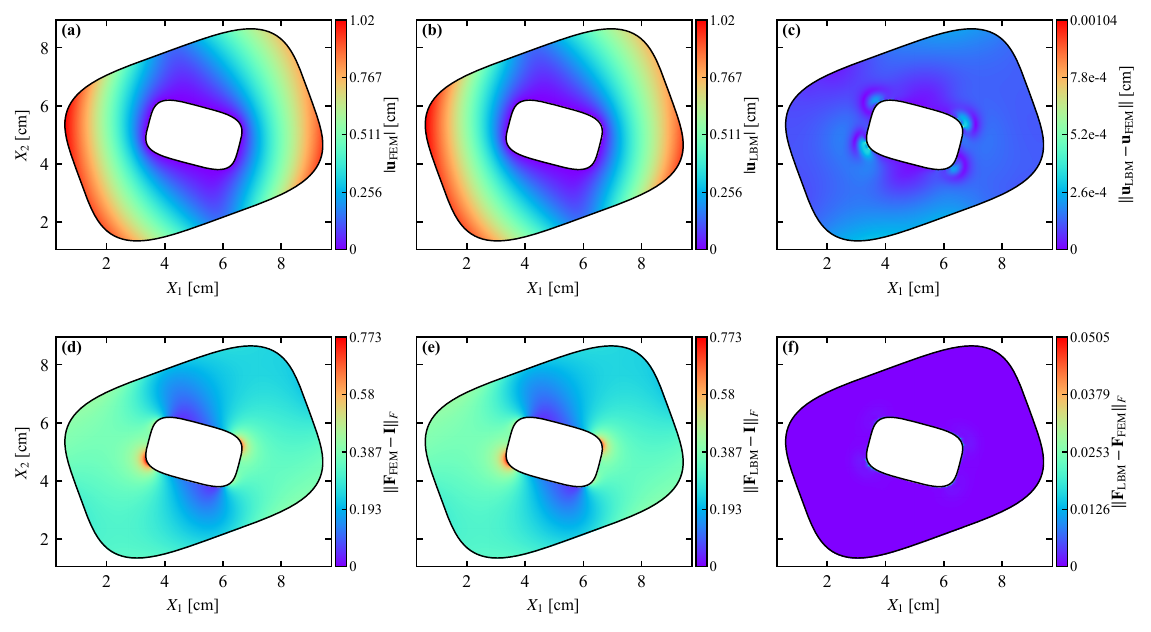}
\caption{Mixed-boundary superellipse shell at \(t=20~{\rm ms}\). The first row compares displacement magnitudes from the finite-element reference and the LBM calculation. The second row compares deformation-gradient increments and the pointwise deformation-gradient difference.}
\label{fig:2d-superellipse-field}
\end{figure}

Figure~\ref{fig:2d-superellipse-field} compares the terminal displacement and deformation-gradient fields on the \(N=384\) lattice.  This grid contains \(60148\) active nodes and \(1676\) cut links.  The maximum displacement magnitude is \(1.0224~{\rm cm}\) in the LBM calculation and \(1.0223~{\rm cm}\) in the finite-element reference.  The relative errors are \(2.88\times10^{-4}\) for displacement, \(5.99\times10^{-4}\) for \(\bF\), and \(2.82\times10^{-3}\) for \(\bF-\Id\).  The terminal Jacobian range is \(0.813\le J\le1.141\), confirming orientation preservation throughout the active domain.  The agreement is significant because the boundary is neither concentric nor aligned with the grid, and the solution has no scalar radial profile that could mask angular boundary errors.

\begin{figure}[pos=htbp]
\centering
\includegraphics[width=0.98\textwidth]{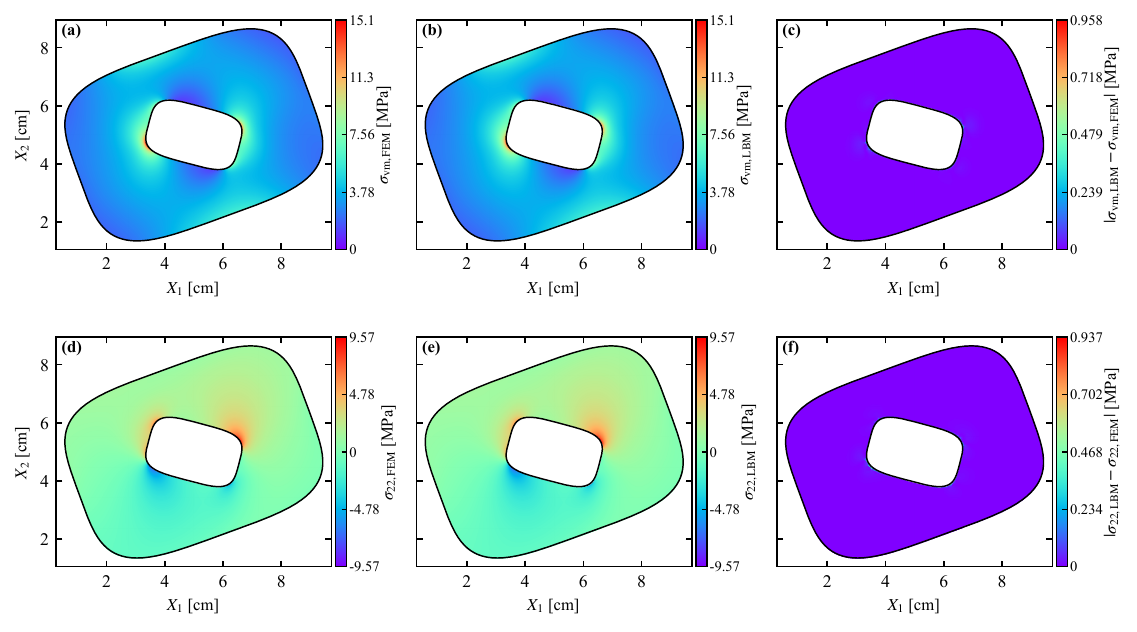}
\caption{Cauchy-stress comparison for the mixed-boundary superellipse shell. The panels report finite-element and LBM von Mises stress, the absolute von Mises difference, the \(\sigma_{22}\) component, and its absolute difference.}
\label{fig:2d-superellipse-stress}
\end{figure}

The stress comparison in Fig.~\ref{fig:2d-superellipse-stress} uses the same recovered Cauchy stress as Eq.~\eqref{eq:cauchy-recovery-general}.  The sampled von Mises stress reaches \(15.03~{\rm MPa}\) in the finite-element reference and \(15.12~{\rm MPa}\) in the LBM calculation.  Both maxima occur at the same grid point adjacent to the inner superellipse, identifying a resolved local stress concentration present in both calculations.  The corresponding \(\sigma_{22}\) ranges are \([-6.91,9.36]~{\rm MPa}\) for FEM and \([-6.91,9.57]~{\rm MPa}\) for LBM.  The largest stress differences occur near the highly curved portions of the inner and outer boundaries.  This is the expected location of the leading error.  There, the boundary deformation gradient is assembled from a surface least-squares derivative and a local nonlinear normal solve.  The interior stress follows directly from the evolved \(\bF\).

\FloatBarrier

\begin{table}[pos=htbp]
\centering
\caption{Summary of the two-dimensional curved-boundary validation cases. Dimensional quantities use the physical scales listed in Section~\ref{subsec:2d-material-scaling}.}
\label{tab:2d-curved-summary}
\scriptsize
\setlength{\tabcolsep}{4pt}
\begin{tabular}{@{}lrl l l p{0.30\textwidth}@{}}
\toprule
Case & \(N\) & Active/cut links & Load and time & Reference & Terminal errors \\
\midrule
Affine annulus & 384 & \(66692/1840\) & \(T=0.2~{\rm ms}\) & analytic & \(E_{L2}(u)=4.55\times10^{-6}\), \(E_{L2}(\bF-\Id)=8.51\times10^{-5}\), \(E_{L2}(\sigma)=5.16\times10^{-5}\) \\
Radial annulus & 196 & \(17396/936\) & \(p_0=10~{\rm MPa}\), \(T=5~{\rm ms}\) & FEM & \(E_{L2}(u)=3.86\times10^{-4}\), \(E_{L2}(\bF)=9.18\times10^{-4}\), \(E_{L2}(\bF-\Id)=3.84\times10^{-3}\) \\
Superellipse shell & 384 & \(60148/1676\) & \(\bar{\bT}=(0,1.25)^T~{\rm MPa}\), \(T=20~{\rm ms}\) & FEM & \(E_{L2}(u)=2.88\times10^{-4}\), \(E_{L2}(\bF)=5.99\times10^{-4}\), \(E_{L2}(\bF-\Id)=2.82\times10^{-3}\) \\
\bottomrule
\end{tabular}
\end{table}

\FloatBarrier

Table~\ref{tab:2d-curved-summary} summarizes the grids, reference solutions, loading conditions, and terminal errors for the two-dimensional validation cases.  Viewed as a set, these cases cover the two boundary parities used by the cut-link reconstruction.  The affine annulus verifies the velocity Dirichlet rule against an exact finite-strain solution in a setting with two curved interfaces.  The radial annulus verifies that the transient lattice dynamics can relax to the same mixed-boundary equilibrium as an independent finite-element reference.  The superellipse shell then probes the same mechanism when symmetry is removed and the traction boundary requires genuinely local normal--tangent reconstruction.  Across the two FEM benchmarks, displacement errors remain at the \(10^{-4}\) level and deformation-gradient errors remain near \(10^{-3}\) for \(\bF\), with stress and gradient errors localized mainly near cut-link boundaries.  This pattern is consistent with the expected accuracy hierarchy.  Primary displacement fields are most accurate, while derivative and stress quantities carry the additional boundary-reconstruction error.

\section{Three-dimensional curved-boundary validation and benchmarks}
\label{sec:3d-curved-validation}

The three-dimensional benchmarks apply the same validation logic to the D3Q6\(\times\)12 formulation, with the same material, physical scaling, and error measures defined in Section~\ref{subsec:2d-material-scaling}.  The extension is substantial because each boundary point now carries two tangent directions, a three-component normal image in the traction inversion, and nine deformation-gradient components in the population state.  The sequence begins with a rotated ellipsoid, where an exact affine field isolates curved velocity Dirichlet reconstruction and gives a clean grid-refinement check.  It then moves to a spherical shell, replacing prescribed motion by curved Neumann data while retaining a nonlinear radial boundary-value problem (BVP) as reference.  The final benchmark removes this radial simplification by using a finite tube that combines planar end constraints, traction-free cylindrical walls, large axial stretch, and finite twist in one transient relaxation problem.

As in Section~\ref{sec:2d-curved-validation}, the equilibrium benchmarks are computed by advancing the time-dependent lattice system.  The terminal comparisons therefore assess whether the D3Q6\(\times\)12 collide--stream dynamics, damping, compatibility projection, and embedded boundaries jointly drive the solution to the correct steady finite-strain state.

\subsection{Rotated ellipsoid with affine Dirichlet data}
\label{subsec:3d-ellipsoid-dirichlet}

The ellipsoid benchmark is selected for the same reason as the affine annulus, with an additional three-dimensional geometric constraint.  The exact solution is affine, so the deformation gradient and stress are spatially uniform; at the same time, the rotated ellipsoidal surface generates cut links whose normals are generally oblique to every lattice direction.  The case therefore probes the three-dimensional Dirichlet rule, the cut-link interpolation, and the active-node mask without introducing interior stress-divergence error.  The material domain is selected by a rotated-ellipsoid level-set mask on the Cartesian lattice,
\begin{equation}
  \Omega_0
  =
  \left\{
    \bX:\ 
    \sum_{\alpha=1}^{3}\left(\frac{\xi_\alpha}{a_\alpha}\right)^2\le1
  \right\},
  \qquad
  \bm\xi=\bm Q^T(\bX-\bX_c),
  \label{eq:ellipsoid-domain-3d}
\end{equation}
where \(\bX_c=(0.5,0.5,0.5)^T\), \((a_1,a_2,a_3)=(0.34,0.25,0.18)\), and \(\bm Q=\bm R_z(25^\circ)\bm R_y(18^\circ)\), with \(\bm R_y\) and \(\bm R_z\) denoting rotations about the \(Y\)- and \(Z\)-axes.  Only nodes satisfying Eq.~\eqref{eq:ellipsoid-domain-3d} are active; the surrounding Cartesian grid provides the background coordinates for locating cut links.  The semi-axes are \(3.4\), \(2.5\), and \(1.8~{\rm cm}\) in physical units, and the centre is located at \((5,5,5)~{\rm cm}\).  This geometry has a single smooth closed boundary.  All cut links on that boundary carry velocity Dirichlet data, and the rotation of the principal axes makes the surface normal vary independently from the Cartesian lattice directions.  The imposed velocity field is affine,
\begin{equation}
  \bv_{\rm ex}(\bX)=\bm A(\bX-\bX_c),
  \qquad
  \bu_{\rm ex}(\bX,t)=t\,\bm A(\bX-\bX_c),
  \label{eq:ellipsoid-affine-solution-3d}
\end{equation}
with
\begin{equation}
  \bm A=
  \begin{pmatrix}
    0 & 0.80 & 0\\
    0 & 0 & -0.60\\
    0.45 & 0 & 0.08
  \end{pmatrix}.
  \label{eq:ellipsoid-A-3d}
\end{equation}
The final time is \(T=0.4599~{\rm ms}\), chosen so that the largest imposed displacement is approximately \(1~{\rm cm}\).  The exact deformation gradient is spatially uniform,
\begin{equation}
  \bF_{\rm ex}(t)=\Id+t\bm A,
  \qquad
  \bF_{\rm ex}(T)=
  \begin{pmatrix}
    1 & 0.3679 & 0\\
    0 & 1 & -0.2759\\
    0.2070 & 0 & 1.0368
  \end{pmatrix},
  \qquad
  J_{\rm ex}(T)=1.01578 .
  \label{eq:ellipsoid-F-3d}
\end{equation}
Since \(\bF_{\rm ex}\) is constant in space, \(\Div\bP(\bF_{\rm ex})=\bzero\).  The calculation therefore examines the curved three-dimensional Dirichlet reconstruction without a bulk forcing term.

\begin{figure}[pos=htbp]
\centering
\includegraphics[width=0.98\textwidth]{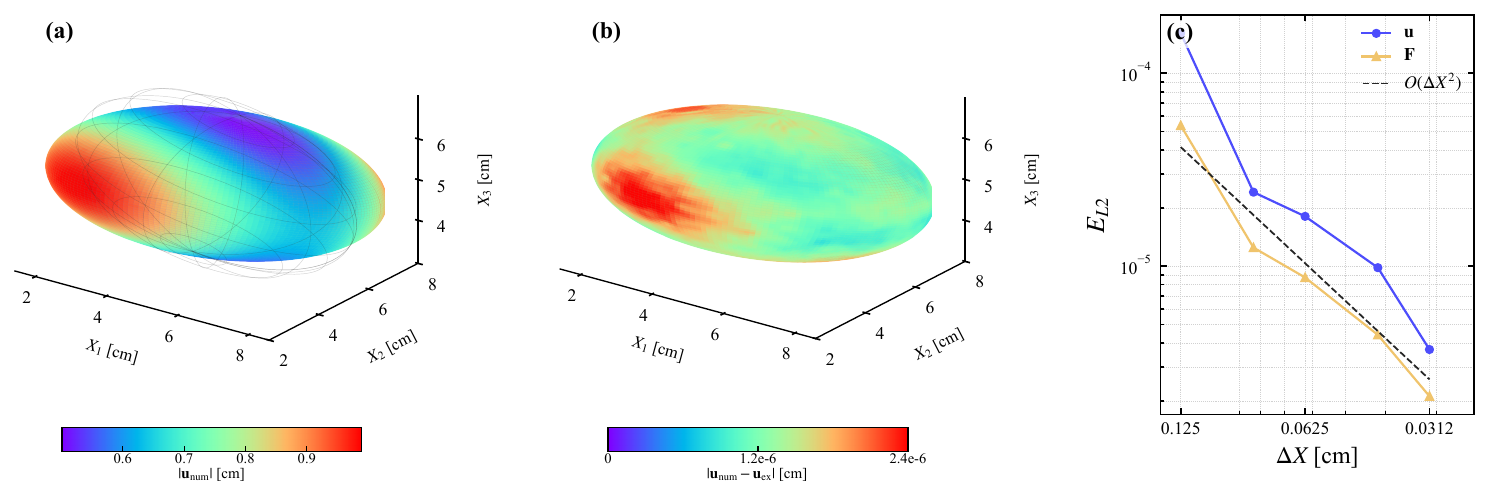}
\caption{Rotated ellipsoid Dirichlet benchmark. Panel (a) shows the exact deformed surface coloured by displacement magnitude, together with the initial wireframe. Panel (b) shows the near-boundary numerical displacement error on the \(N=320\) grid. Panel (c) reports grid convergence for displacement and deformation gradient.}
\label{fig:3d-ellipsoid-dirichlet}
\end{figure}

Figure~\ref{fig:3d-ellipsoid-dirichlet} shows the deformed ellipsoid, the near-boundary displacement error, and the grid refinement from \(N=80\) to \(320\).  The finest grid has \(\Delta x=0.03125~{\rm cm}\), \(2.10\times10^6\) active nodes, and \(1.25\times10^5\) cut links.  At this resolution the relative errors are \(3.70\times10^{-6}\) for displacement, \(2.15\times10^{-5}\) for velocity, and \(2.12\times10^{-6}\) for \(\bF\).  The terminal Jacobian lies in the interval \(1.01577\le J\le1.01578\), matching the spatially constant exact value in Eq.~\eqref{eq:ellipsoid-F-3d}.

The refinement curve has small departures from a perfectly straight power law.  This behaviour is expected for an embedded ellipsoidal boundary because the active-node set, cut-link locations, and near-boundary quadrature points change discretely with resolution; the comparison therefore lacks a nested sequence of identical boundary samples.  Even with this discrete sampling effect, both \(u\) and \(\bF\) remain close to the second-order guide and reach relative errors of order \(10^{-6}\) on the finest grid.  The ellipsoid benchmark thus supports near-second-order behaviour for the three-dimensional curved Dirichlet reconstruction despite the non-boundary-fitted surface.

\FloatBarrier

\subsection{Spherical shell radial traction benchmark}
\label{subsec:3d-spherical-shell}

The spherical shell is the three-dimensional radial-traction analogue of the annulus with a distinct boundary setting.  The annulus benchmark is mixed Dirichlet--Neumann, with a clamped inner boundary and a loaded outer boundary.  The spherical-shell lattice calculation treats both curved surfaces as Neumann boundaries.  Both boundary components are curved surfaces, the inner and outer normals have opposite radial signs, and the Neumann rule resolves a three-component normal image of \(\bF\) at each cut link.  The radial reduction supplies an exact nonlinear boundary-value problem for the terminal state.  The lattice calculation still follows a transient ramp-and-relax path.  The material is the spherical layer between an inner cavity and an outer spherical surface,
\begin{equation}
  \Omega_0
  =
  \left\{\bX:\ R_i\le R\le R_o\right\},
  \qquad
  R=\|\bX-\bX_c\|,
  \qquad
  R_i=0.20,\quad R_o=0.38 .
  \label{eq:spherical-shell-domain-3d}
\end{equation}
Here \(\bX_c=(0.5,0.5,0.5)^T\).  The physical inner and outer radii are \(2.0~{\rm cm}\) and \(3.8~{\rm cm}\), giving a wall thickness of \(1.8~{\rm cm}\).  The level-set representation treats the two spherical surfaces as separate boundary components.  On the outer sphere the outward normal is \(\be_R\); on the inner cavity surface it is \(-\be_R\).  This sign distinction is used by the boundary identifiers that assign the inner and outer Neumann data.  A radially symmetric reference solution uses the same radial ansatz as Eq.~\eqref{eq:radial-map}, \(\bx(\bX)=\bX_c+r(R)\be_R\), with \(R=\|\bX-\bX_c\|\) and \(\be_R=(\bX-\bX_c)/R\).  The three-dimensional difference from the annular reduction is the duplicated tangential stretch,
\begin{equation}
  \lambda_r=\frac{\dd r}{\dd R},
  \qquad
  \lambda_\theta=\lambda_\phi=\frac{r}{R},
  \qquad
  J=\lambda_r\lambda_\theta^2 .
  \label{eq:spherical-stretches-3d}
\end{equation}
Following the standard radial reduction for finite elasticity \citep{ogden1997,holzapfel2000}, the same neo-Hookean principal-stress relation as Eq.~\eqref{eq:radial-principal-stresses-2d}, evaluated with the stretches in Eq.~\eqref{eq:spherical-stretches-3d}, gives the spherical equilibrium equation
\begin{equation}
  \frac{\dd P_r}{\dd R}
  +\frac{2(P_r-P_\theta)}{R}=0,
  \qquad
  r(R_i)=0.28,
  \qquad
  P_r(R_o)=0 .
  \label{eq:spherical-equilibrium-3d}
\end{equation}
Solving Eq.~\eqref{eq:spherical-equilibrium-3d} gives \(r(R_o)=0.42331\), \(P_r(R_i)=-1.15242\), \(0.5526\le\lambda_r\le0.9387\), \(1.114\le\lambda_\theta\le1.400\), and \(1.083\le J\le1.165\).  The condition \(r(R_i)=0.28\) is used to determine the pressure level of the radial reference state; the lattice calculation does not impose this displacement.  Instead, both shell surfaces are treated as Neumann boundaries.  The outer surface is traction-free, and the inner surface receives the corresponding ramped normal traction; in physical units this expands the cavity with an amplitude of \(11.52~{\rm MPa}\).  The load uses the same sine-squared ramp as Eq.~\eqref{eq:sine2-ramp}, with \(T_r=2~{\rm ms}\), and the run is advanced to \(5~{\rm ms}\) with damping and compatibility projection.

\begin{figure}[pos=htbp]
\centering
\includegraphics[width=0.98\textwidth]{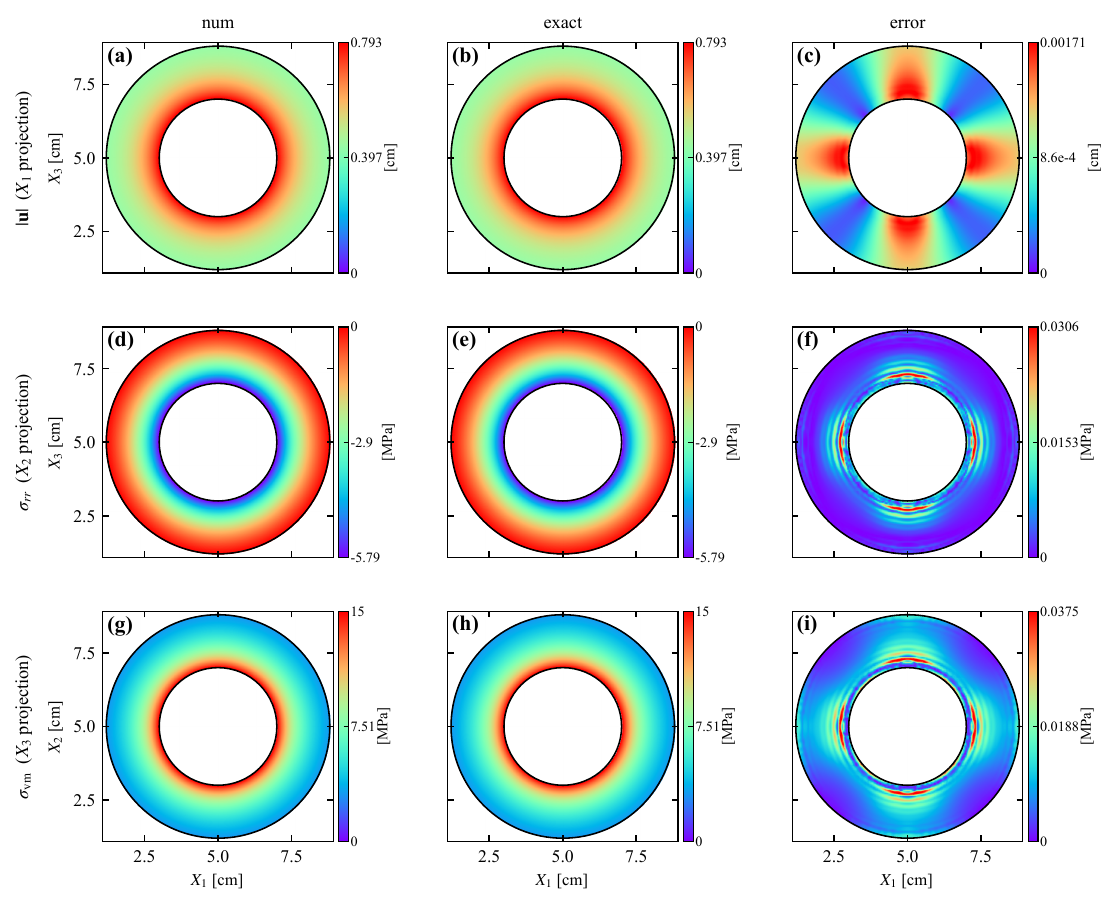}
\caption{Spherical shell radial traction benchmark at \(t=5~{\rm ms}\). The rows show displacement magnitude, radial Cauchy stress \(\sigma_{rr}\), and von Mises stress. The columns compare the LBM solution, the exact radial BVP solution, and the absolute error on representative cut planes.}
\label{fig:3d-spherical-shell-fields}
\end{figure}

Figure~\ref{fig:3d-spherical-shell-fields} compares the LBM and exact fields on the \(N=200\) grid.  The grid spacing is \(0.05~{\rm cm}\), with \(1.57\times10^6\) active nodes and \(1.39\times10^5\) cut links.  The relative errors are \(1.18\times10^{-3}\) for displacement, \(5.23\times10^{-4}\) for \(\bF\), and \(2.07\times10^{-3}\) for Cauchy stress.  The residual velocity has root-mean-square (RMS) value \(4.56\times10^{-4}\), and the numerical Jacobian remains in the range \(1.081\le J\le1.165\).  Together, these diagnostics show that the reported field is a relaxed terminal equilibrium of the transient lattice dynamics after the pressure ramp.

\begin{figure}[pos=htbp]
\centering
\includegraphics[width=0.98\textwidth]{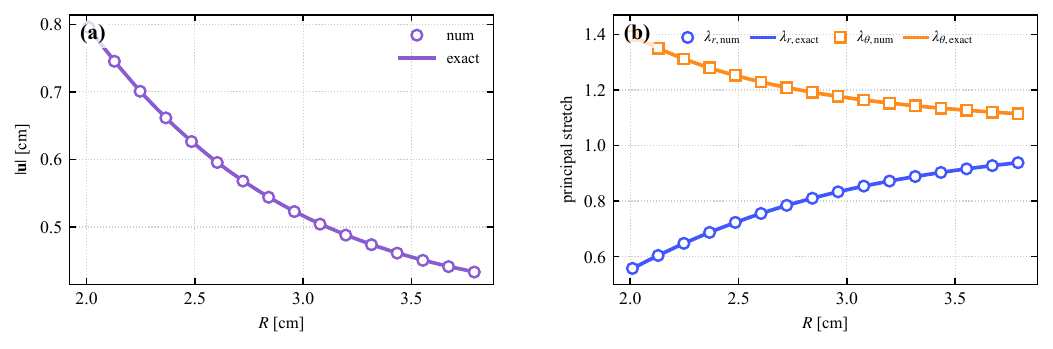}
\caption{Radial profiles for the spherical shell benchmark. Panel (a) compares displacement magnitude. Panel (b) compares the radial stretch \(\lambda_r\) and tangential stretch \(\lambda_\theta\) from Eq.~\eqref{eq:spherical-stretches-3d}.}
\label{fig:3d-spherical-shell-profiles}
\end{figure}

Figure~\ref{fig:3d-spherical-shell-profiles} gives the radial profiles extracted from the same terminal field.  Panel (a) plots the displacement magnitude against the reference radius, showing the decay from the expanded inner surface to the traction-free outer surface.  Panel (b) separates the principal stretches.  The radial stretch increases from about \(0.55\) near the cavity to about \(0.94\) at the exterior surface, while the tangential stretch decreases from about \(1.40\) to about \(1.11\).  In both panels, the open markers are radial-bin averages of the LBM data and the solid curves are the exact BVP profiles.  These profiles also check the principal stretch decomposition that enters the nonlinear nominal traction condition.

\FloatBarrier

\subsection{Finite tube under end stretch and twist}
\label{subsec:3d-tube-pull-torsion}

\begin{figure}[pos=htbp]
\centering
\includegraphics[width=0.98\textwidth]{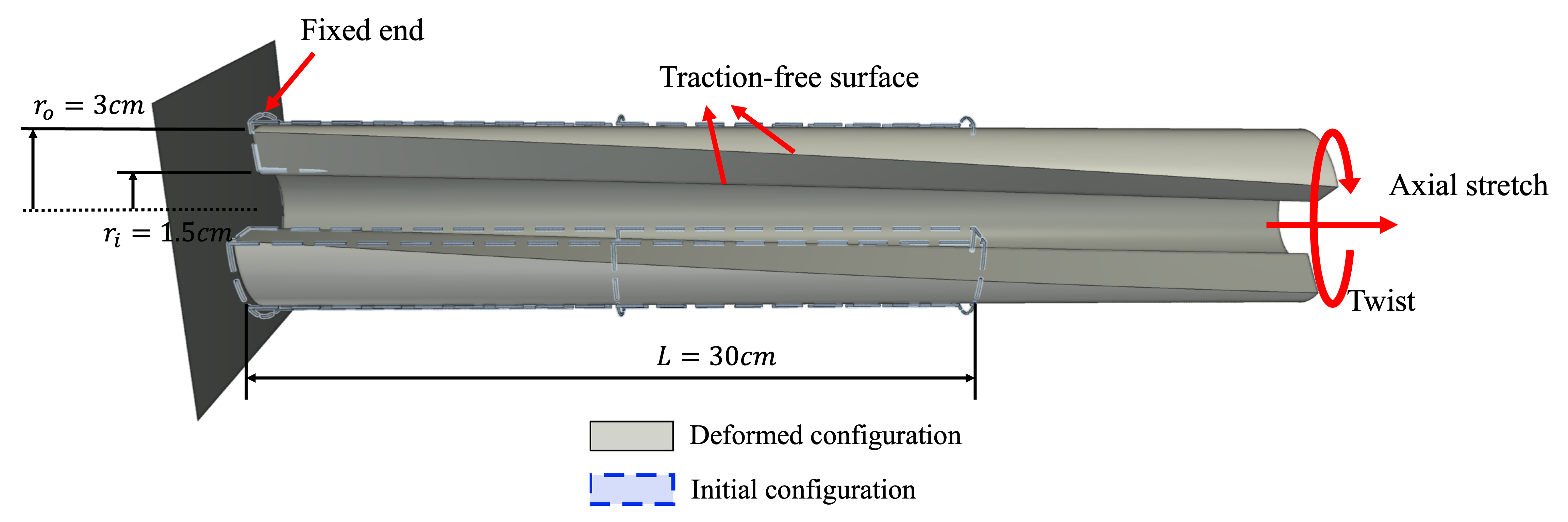}
\caption{Schematic of the finite tube benchmark. The left end is fixed, the cylindrical walls are traction-free, and the right end is loaded by prescribed axial stretch and twist. The dashed surface indicates the undeformed reference configuration.}
\label{fig:3d-tube-schematic}
\end{figure}

The final benchmark is designed to evaluate the algorithm in a geometry closer to finite-strain structural computation than to a manufactured solution.  It combines several boundary identifiers on one embedded body, includes junctions between planar end faces and curved cylindrical walls, and imposes a large terminal displacement through simultaneous axial extension and twist.  The cylindrical walls are traction-free Neumann boundaries, so the benchmark also checks whether the local traction reconstruction remains stable during a long relaxation from a highly deformed configuration.  The reference body is a finite hollow cylinder whose axis is the \(Z\)-direction.  In each cross-section the material is an annulus, and along the axis this annulus is extruded over a finite length.  The domain is
\begin{equation}
  \Omega_0
  =
  \left\{
    \bX:\ R_i\le R\le R_o,\ 0\le Z\le L
  \right\},
  \qquad
  R=\sqrt{(X_1-X_c)^2+(X_2-Y_c)^2},
  \label{eq:tube-domain-3d}
\end{equation}
where \(Z=X_3\).  The transverse centre is \((X_c,Y_c)=(0.38,0.38)\) in the computational box.  With \(R_i=0.15\), \(R_o=0.30\), and \(L=3.0\), the physical dimensions are \(R_i=1.5~{\rm cm}\), \(R_o=3.0~{\rm cm}\), and \(L=30~{\rm cm}\); the wall thickness is \(1.5~{\rm cm}\) and the aspect ratio is \(L/R_o=10\).  The boundary has four components, namely the outer cylindrical wall \(R=R_o\), the inner cylindrical wall \(R=R_i\), the fixed annular end face \(Z=0\), and the loaded annular end face \(Z=L\).  The two cylindrical walls are traction-free curved Neumann boundaries.  The end faces are treated as planar Dirichlet components within the same boundary-component labelling scheme.  With \(\bX_\perp=(X_1-X_c,X_2-Y_c)^T\), \(\bm R_\alpha\) denoting the \(2\times2\) rotation matrix through angle \(\alpha\), and \(\Id_2\) denoting the \(2\times2\) identity, the right-end displacement is
\begin{equation}
  \bu_D(\bX,t)
  =
  \begin{pmatrix}
    \left(\bm R_{\alpha(t)}-\Id_2\right)\bX_\perp\\
    \Delta L(t)
  \end{pmatrix},
  \qquad
  \alpha(t)=s(t)\Phi,
  \qquad
  \Delta L(t)=s(t)\varepsilon_z L ,
  \label{eq:tube-right-displacement}
\end{equation}
where \(\Phi=45^\circ\), \(\varepsilon_z=0.50\), and
\begin{equation}
  s(t)=
  \begin{cases}
    \dfrac{1-\cos(\pi t/T_r)}{2}, & 0\le t<T_r,\\[5pt]
    1, & t\ge T_r,
  \end{cases}
  \qquad
  T_r=10~{\rm ms}.
  \label{eq:tube-cosine-ramp}
\end{equation}
The solution is then relaxed to \(t=100~{\rm ms}\).  Figure~\ref{fig:3d-tube-schematic} shows the boundary layout used for the calculation.

The reported three-dimensional field is computed on the finest embedded lattice, with \(76\times76\times300\) nodes, \(\Delta x=0.10~{\rm cm}\), \(6.34\times10^5\) active nodes, and \(1.12\times10^5\) cut links.  The cut links are distributed over the outer wall, inner wall, and the two end faces, so this case exercises both curved Neumann reconstruction and planar Dirichlet reconstruction in a single run.  The terminal maximum displacement is \(15.14~{\rm cm}\), and the Jacobian stays positive with \(1.236\le J\le1.547\).  The positive Jacobian range is a useful stability diagnostic because the imposed deformation combines a \(50\%\) axial stretch with a \(45^\circ\) twist over a long tube; loss of compatibility or boundary inconsistency would first appear as localized distortion near the end--wall junctions or the traction-free surfaces.

\begin{figure}[pos=htbp]
\centering
\includegraphics[width=0.98\textwidth]{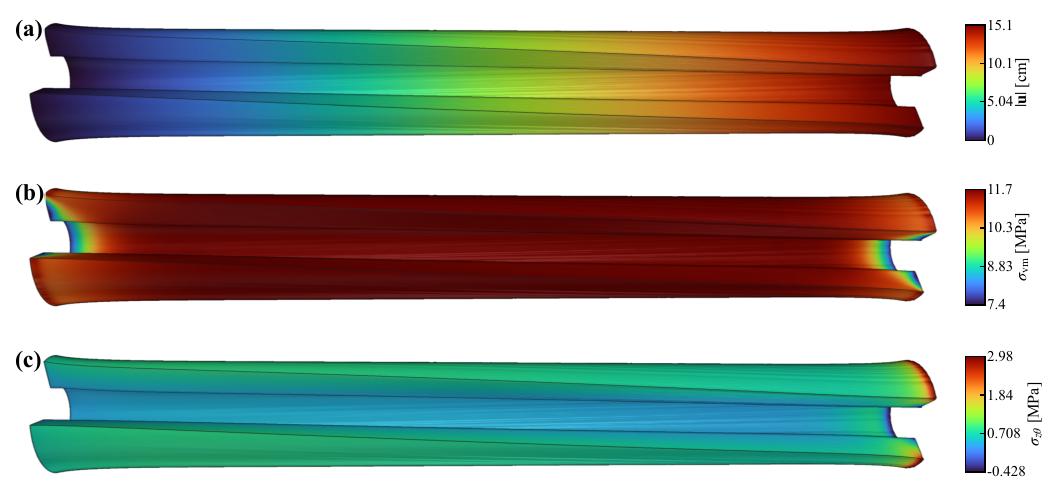}
\caption{Finite tube under \(50\%\) axial stretch and \(45^\circ\) twist at \(t=100~{\rm ms}\). The cutaway surfaces show (a) displacement magnitude, (b) Cauchy von Mises stress, and (c) the axial-hoop shear stress \(\sigma_{z\theta}\).}
\label{fig:3d-tube-fields}
\end{figure}

\FloatBarrier

\begin{figure}[pos=htbp]
\centering
\includegraphics[width=0.98\textwidth]{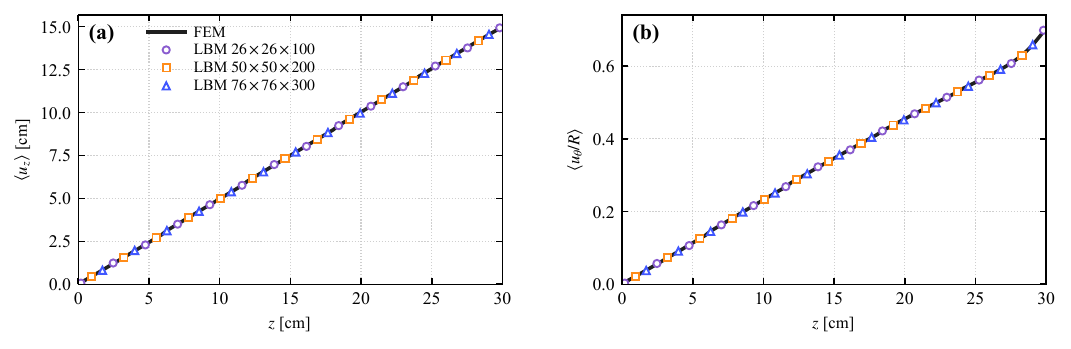}
\caption{Axial profile comparison for the finite tube benchmark at \(t=100~{\rm ms}\). The solid line is the finite-element reference, and the open markers show LBM results on \(26\times26\times100\), \(50\times50\times200\), and \(76\times76\times300\) grids. Panel (a) compares the cross-section mean axial displacement. Panel (b) compares the cross-section mean twist measure \(u_\theta/R\).}
\label{fig:3d-tube-profiles}
\end{figure}

An independent finite-element calculation \citep{bonetwood2008} on a boundary-fitted hexahedral tube mesh provides the quantitative reference.  The updated axial-profile comparison uses three embedded lattices, \(26\times26\times100\), \(50\times50\times200\), and \(76\times76\times300\), with physical spacings \(0.30\), \(0.15\), and \(0.10~{\rm cm}\).  Because the LBM and finite-element discretizations use different volumetric meshes, the most robust quantitative comparison is made through cross-section averages along the tube axis.  These profiles retain the axial extension, torsional rotation, and volumetric response while suppressing interpolation noise from local mesh-to-lattice sampling.  For a scalar field \(q\), the axial profile comparison uses
\begin{equation}
  \langle q\rangle(Z)=\frac{1}{|\Omega_Z|}\int_{\Omega_Z}q\,\dd A,
  \qquad
  \frac{u_\theta}{R}
  =
  \frac{-(X_2-Y_c)u_1+(X_1-X_c)u_2}{R^2}.
  \label{eq:tube-profile-definitions}
\end{equation}
Here \(\Omega_Z\) is the cross-section of \(\Omega_0\) at axial coordinate \(Z\), and \(u_z=u_3\) is the axial displacement component.  Figure~\ref{fig:3d-tube-profiles} compares \(\langle u_z\rangle\) and \(\langle u_\theta/R\rangle\) against the finite-element profiles.  The finite-element result is plotted as a solid curve, and the LBM results are plotted as open markers at the three resolutions.  All three embedded-lattice profiles follow the same axial stretch and twist trends at the scale of the figure.  On the finest grid, the terminal root-mean-square error (RMSE) is \(1.90\times10^{-3}\) in nondimensional units for \(\langle u_z\rangle\), corresponding to \(1.90\times10^{-2}~{\rm cm}\), and \(8.46\times10^{-4}\) for \(\langle u_\theta/R\rangle\).  The corresponding \(J\)-profile RMSE is \(4.87\times10^{-3}\).  The \(50\times50\times200\) grid gives similar averaged-profile errors, with \(1.32\times10^{-2}~{\rm cm}\) for \(\langle u_z\rangle\), \(5.84\times10^{-4}\) for \(\langle u_\theta/R\rangle\), and \(4.96\times10^{-3}\) for \(J\).  These averaged profiles provide a resolution-consistency check for the relaxed tube deformation by showing that the axial stretch, twist, and volumetric response remain stable across the three embedded lattices.

\FloatBarrier

\begin{table}[pos=htbp]
\centering
\caption{Summary of the three-dimensional curved-boundary validation cases. Dimensional quantities use the physical scales listed in Section~\ref{subsec:2d-material-scaling}.}
\label{tab:3d-curved-summary}
\scriptsize
\setlength{\tabcolsep}{4pt}
\begin{tabular}{@{}l l r l p{0.36\textwidth}@{}}
\toprule
Case & Grid & Active/cut links & Reference & Terminal measures \\
\midrule
Rotated ellipsoid & \(320^3\) & \(2.10{\times}10^6/1.25{\times}10^5\) & analytic & \(T=0.460~{\rm ms}\), \(E_{L2}(u)=3.70\times10^{-6}\), \(E_{L2}(v)=2.15\times10^{-5}\), \(E_{L2}(\bF)=2.12\times10^{-6}\) \\
Spherical shell & \(200^3\) & \(1.57{\times}10^6/1.39{\times}10^5\) & analytic BVP & \(T=5~{\rm ms}\), \(E_{L2}(u)=1.18\times10^{-3}\), \(E_{L2}(\bF)=5.23\times10^{-4}\), \(E_{L2}(\sigma)=2.07\times10^{-3}\) \\
Tube pull--torsion & \(76{\times}76{\times}300\) & \(6.34{\times}10^5/1.12{\times}10^5\) & FEM & \(T=100~{\rm ms}\), \({\rm RMSE}(\langle u_z\rangle)=1.90\times10^{-3}\), \({\rm RMSE}(\langle u_\theta/R\rangle)=8.46\times10^{-4}\), \({\rm RMSE}(\langle J\rangle)=4.87\times10^{-3}\) \\
\bottomrule
\end{tabular}
\end{table}

\FloatBarrier

Table~\ref{tab:3d-curved-summary} collects the grid sizes, reference solutions, and terminal measures for the three-dimensional sequence.  The results show that the reconstruction logic carries over from D2Q4\(\times\)6 to D3Q6\(\times\)12 after adding full surface geometry and the nine deformation-gradient components.  The ellipsoid demonstrates near-second-order convergence for a non-axis-aligned curved Dirichlet boundary, with the expected small irregularities caused by non-nested embedded-boundary sampling.  The spherical shell verifies the curved Neumann reconstruction against a nonlinear radial BVP and shows that the transient dynamics relax to a terminal state with small residual velocity.  The finite tube then combines end Dirichlet conditions, traction-free cylindrical walls, compatibility projection, and finite-element profile agreement on a large embedded domain.  Alongside the two-dimensional benchmarks, these results support the intended use of the method as a transient LBM solver for curved-domain finite-strain hyperelasticity whose steady benchmarks are recovered as relaxed long-time limits.

\section{Conclusions and outlook}
\label{sec:conclusions-outlook}

This work has developed a total-Lagrangian vectorial lattice Boltzmann method for finite-strain hyperelastic dynamics on curved embedded domains.  The conservative state is formed by the material velocity and the full deformation gradient, and vector-valued lattice populations represent both the state and the material-coordinate fluxes.  The same dimensional construction gives the D2Q4\(\times\)6 scheme in two dimensions and the D3Q6\(\times\)12 scheme in three dimensions.  The nonlinear material response enters the bulk update through the local Piola-stress map \(\bP(\bF)\), so the method retains explicit collide--stream dynamics on Cartesian lattices in a total-Lagrangian solid formulation.

Curved geometry enters the algorithm at the cut links, where the Cartesian lattice loses the incoming population after streaming.  The level-set description supplies the boundary point, normal, and cut fraction, and the opposite-population identities decide row by row whether a state quantity or a link flux is being imposed.  Velocity and nominal-traction boundaries are handled within this same reconstruction framework.  The velocity case uses the prescribed material motion and its kinematic flux, and the traction case builds a compatible boundary deformation gradient from surface derivatives and a local nonlinear solve.  The compatibility projection keeps the recovered displacement aligned with the evolved deformation gradient during traction-boundary updates, allowing curved Dirichlet, Neumann, and mixed boundaries to be imposed on the embedded Cartesian lattice.

The validation sequence follows the same progression in geometry and boundary complexity.  The two-dimensional annulus benchmarks first separate exact curved velocity reconstruction from mixed traction loading, and the superellipse then removes circular symmetry through changing normals and wall thickness.  The three-dimensional cases repeat this logic at larger geometric cost, with the ellipsoid providing an exact affine benchmark for D3Q6\(\times\)12, the spherical shell exercising curved Neumann reconstruction against a nonlinear radial solution, and the stretched--twisted tube combining end constraints with traction-free curved walls.  Displacement fields agree closely with exact or finite-element references throughout the sequence.  The larger errors in deformation-gradient and stress quantities remain mainly near embedded cut-link boundaries, consistent with their derivative and constitutive character.

For extensions beyond the neo-Hookean examples considered here, the useful feature is that the constitutive law remains a local material-point operation.  Anisotropic hyperelasticity can be incorporated by replacing the strain-energy density and its Piola derivative without changing the lattice moment structure.  Thermoelasticity can be introduced through a temperature-dependent energy, thermal distortion, or eigenstrain-type expansion, with temperature advanced by a companion transport solver.  Rate-dependent and history-dependent models follow the same pattern.  Viscoelastic branches, plastic deformation gradients, hardening variables, or overstress variables would be updated locally at each lattice node, and the resulting Piola map would feed the same bulk update and traction-boundary solve.

This locality is also the reason the method is well matched to high-performance material mechanics.  Regular memory access, nearest-neighbor communication, and local constitutive evaluation suit graphics processing units (GPUs) and massively parallel architectures, and the embedded Cartesian lattice avoids mesh distortion and repeated remeshing under large body motion.  Important challenges remain, including near-incompressibility, strong anisotropy, plastic localization, contact, fracture, and robust traction inversion under severe nonlinear response.  The shared population-based structure also suggests a path toward unified multiphysics solvers in which solid, fluid, thermal, and mass-transport LBM components use a common computational architecture.  Such coupling would be useful for soft materials, biological tissues, porous media, architected materials, and manufacturing processes where deformation, flow, heat transfer, and diffusion interact.

Overall, the present work extends lattice Boltzmann methodology toward curved-domain finite-strain solid mechanics.  By combining a total-Lagrangian first-order solid formulation, vector-valued moment matching, level-set cut-link geometry, and compatible boundary reconstruction, the method provides an explicit and geometry-flexible route for nonlinear elastic solids on Cartesian lattices.

\section*{Data availability}

All lattice Boltzmann simulations reported in this article were carried out on an NVIDIA GeForce RTX 5090 graphics processing unit.  The data underlying this article are available in Zenodo at \href{https://doi.org/10.5281/zenodo.20572218}{https://doi.org/10.5281/zenodo.20572218}.

\clearpage
\appendix
\numberwithin{equation}{section}
\renewcommand{\thesection}{Appendix~\Alph{section}}
\renewcommand{\thesubsection}{\Alph{section}.\arabic{subsection}}
\renewcommand{\theequation}{\Alph{section}.\arabic{equation}}
\section{Grid-aligned D3Q6\texorpdfstring{\(\times\)}{x}12 boundary rules}
\label{app:grid-aligned-d3q6-boundaries}

This appendix records the half-way boundary formulas for a rectangular three-dimensional domain.  They are the grid-aligned limit of the curved reconstruction in Section~\ref{sec:curved-boundary-reconstruction} and are useful for verification.  The state ordering is Eq.~\eqref{eq:state-3d}, and the D3Q6 directions are \(\be_q\in\{\pm\be_1,\pm\be_2,\pm\be_3\}\).

\subsection{Pair identities}
\label{appsubsec:pair-ids-3d}

For a face with outward normal \(\bN=\pm\be_A\), the missing incoming direction is \(\bd=-\bN\).  The D3Q6 equilibrium gives
\begin{equation}
  \bm f_d^{eq}+\bm f_{-d}^{eq}=\frac13\bU,
  \qquad
  \bm f_d^{eq}-\bm f_{-d}^{eq}=\frac{1}{c}d_A\bPhi_A(\bU).
  \label{eq:appendix-pair-3d}
\end{equation}
At a half-way boundary, the reconstruction is
\begin{equation}
  \bm f_d^{n+1}
  =
  \bm D\,\bm f_{-d}^{*,n}
  +\bm S^{n+1/2} .
  \label{eq:appendix-halfway-3d}
\end{equation}
where \(\bm D\) and \(\bm S\) take the Dirichlet or Neumann values given below.  All signs in the following formulas are direct consequences of Eq.~\eqref{eq:appendix-pair-3d}.

\subsection{Velocity Dirichlet boundary}
\label{appsubsec:dirichlet-3d}

For a prescribed displacement boundary, the imposed first-order boundary value is \(\bv_D=\partial_t\bu_D\).  The velocity rows impose a state value; the deformation-gradient rows impose the kinematic normal flux.  Hence
\begin{equation}
  \bm D_D=\diag(-I_3,I_9),
  \label{eq:appendix-DD-3d}
\end{equation}
with
\begin{equation}
  \bm S_D=
  \left(
  \frac{v_{D,1}}{3},\frac{v_{D,2}}{3},\frac{v_{D,3}}{3},
  \frac{N_1v_{D,1}}{c},\frac{N_2v_{D,1}}{c},\frac{N_3v_{D,1}}{c},
  \frac{N_1v_{D,2}}{c},\frac{N_2v_{D,2}}{c},\frac{N_3v_{D,2}}{c},
  \frac{N_1v_{D,3}}{c},\frac{N_2v_{D,3}}{c},\frac{N_3v_{D,3}}{c}
  \right)^T .
  \label{eq:appendix-SD-3d}
\end{equation}
The identity \(\bd=-\bN\) converts the curved-link expression \(S_D^{F_{iB}}=-d_Bv_{D,i}/c\) into the grid-aligned form above.

\subsection{Nominal-traction Neumann boundary}
\label{appsubsec:neumann-3d}

For a prescribed nominal traction \(\bar{\bT}=\bP\bN\), the velocity rows impose the stress flux and the deformation-gradient rows impose a boundary state.  The parity matrix is
\begin{equation}
  \bm D_N=\diag(I_3,-I_9),
  \label{eq:appendix-DN-3d}
\end{equation}
with
\begin{equation}
  \bm S_N=
  \left(
  \frac{\bar T_1}{c},\frac{\bar T_2}{c},\frac{\bar T_3}{c},
  \frac{F_{11}^b}{3},\frac{F_{12}^b}{3},\frac{F_{13}^b}{3},
  \frac{F_{21}^b}{3},\frac{F_{22}^b}{3},\frac{F_{23}^b}{3},
  \frac{F_{31}^b}{3},\frac{F_{32}^b}{3},\frac{F_{33}^b}{3}
  \right)^T .
  \label{eq:appendix-SN-3d}
\end{equation}
If \(\bN=\operatorname{sign}(N_A)\be_A\) for the unique coordinate \(A\) normal to the face, the two tangential columns of \(\bF^b\) are extrapolated or reconstructed from interior data.  The normal column \(\bF^b_{:A}\in\RR^3\), where \(:A\) denotes column \(A\), is determined by
\begin{equation}
  \operatorname{sign}(N_A)P_{iA}(\bF^b)=\bar T_i,
  \qquad i=1,2,3 .
  \label{eq:appendix-axis-newton}
\end{equation}
This is a three-unknown constitutive inversion at each boundary link.  Newton iteration with line search is used, and trial states with \(\det\bF^b\le0\) are rejected.  The calculation is independent from one boundary link to another, preserving the local update pattern of the lattice method.

\end{document}